\documentclass[epj,final]{svjour}
\usepackage[dvips]{graphicx}
\usepackage{hyperref}

\begin{document}

\title{Wave function of the Universe in the early stage of its evolution}


\author{Sergei~P.~Maydanyuk%
\thanks{\emph{E-mail:} maidan@kinr.kiev.ua}}
\offprints{Sergei~P.~Maydanyuk}
\institute{Institute for Nuclear Research, National Academy of Sciences of Ukraine, \\
47, prosp. Nauki, Kiev-28, 03680, Ukraine}
\date{\small\today}
%
\abstract{In quantum cosmological models, constructed in the framework of Friedmann--Robertson--Walker metrics, a nucleation of the Universe with its further expansion is described as a tunneling transition through an effective barrier between regions with small and large values of the scale factor $a$ at non-zero (or zero) energy.
The approach for describing this tunneling consists of constructing a wave function satisfying an appropriate boundary condition. There are various ways for defining the boundary condition that lead to different estimates of the barrier penetrability and the tunneling time. \newline
In order to describe the escape from the tunneling region as accurately as possible and to construct the total wave function on the basis of its two partial solutions unambiguously, we use the tunneling boundary condition that \emph{the total wave function must represent only the outgoing wave at the point of escape from the barrier}, where the following definition for the wave is introduced: \emph{the wave is represented by the wave function whose modulus changes minimally under a variation of the scale factor $a$}.
We construct a new method for a direct non-semiclassical calculation of the total stationary wave function of the Universe, analyze the behavior of this wave function in the tunneling region, near the escape point and in the asymptotic region, and estimate the barrier penetrability.
\emph{We observe oscillations of modulus of wave function in the external region starting from the turning point which decrease with increasing of $a$ and which are not shown in semiclassical calculations.}
The period of such an oscillation decreases uniformly with increasing $a$ and can be used as a fully quantum dynamical characteristic of the expansion of the Universe.
}
\maketitle
\hugehead

\section{Introduction}

Tracking the evolution of the Universe back in time on the basis of the classical equations of general relativity, one encounters singularities, where such equations break down. In order to understand what really happens in the formation of the Universe, many people came to the point of view that a quantum consideration of this process is the deeper one.
So, according to~\cite{Vilenkin.1995}, in the quantum approach we have the following picture of the Universe creation: a closed Universe with a small size is formed from ``nothing'' (vacuum), where by the word ``nothing'' one refers to a quantum state without classical space and time. Here, the wave function can be used for a probabilistic description of the creation of the Universe and its subsequent expansion.
The first papers with the quantum approach for the description of Universe formation and its initial expansion may be~\cite{DeWitt.1967,Wheeler.1968}, and shortly afterwards many other papers appeared in this field, pointing to a rapid development of the quantum approach in cosmology (for example, see \cite{Vilenkin.1982.PLB,Hartle.1983.PRD,Linde.1984.LNC,Zeldovich.1984.LNC,Rubakov.1984.PLB,Vilenkin.1984.PRD,Vilenkin.1986.PRD,Fomin.1975.DAN,Atkatz.1984.PRD} and some discussions in~\cite{Vilenkin.1994.PRD,Rubakov.1999} and references therein).

A key point in the construction of the proper wave function is the choice of a boundary condition. We find a natural and clear condition in the papers of Vilenkin --- \emph{the tunneling boundary condition} \cite{Vilenkin.1995,Vilenkin.1994.PRD}, according to which the wave function in the region of the escape point from the barrier must represent only an outgoing wave. The simplest answer to this question is given in the semiclassical approach (for example, see Refs.~\cite{Vilenkin.1994.PRD,Rubakov.1999}).
However, the validity of the semiclassical approximation decreases when we pass closer to the escape point, which is exactly the point where we want to impose the boundary condition, and we must take into account exact solutions here.
It is natural to suppose that the direct quantum approach is richer, accurate and detailed. And perhaps, it should be more effective going beyond the semiclassical study of nucleation of the Universe and its further evolution.
On the other side, the semiclassical methods are practically shown as accurate enough in a quantum calculation of the rates and other characteristics of the evolution of the Universe in cosmological models.
The semiclassical methods are more stable and convergent in computer calculations (which is very important for obtaining reliable values of the calculated parameters in cosmological scales). At present, a semiclassical power formalism has been developed where the physical meaning of the parameters and characteristics is clear enough intuitively \cite{Ambjorn.2005.PLB}.
On such a basis, one can ask the question: \emph{whether is the direct non-semiclassical quantum approach for the determination of the wave function of the Universe really better in comparison with semiclassical methods?}

Let us find such characteristics, which are less effectively determined by the semiclassical approach.
In order to do this, let us consider the rules of correspondence between two semiclassical wave functions in the tunneling region and in the above barrier region close to the turning point up to the second correction of the semiclassical approximation (for example, we use (47,5) in p.~208 and (50,2) in p.~221 of~\cite{Landau.v3.1989}).
We conclude to the following.
\begin{itemize}

\item
The semiclassical wave function thus defined in the tunneling region is real and we obtain a zero flux in this region. In other words, we have no any propagation of waves under the barrier. So, the semiclassical approximation is less effective for study of tunneling processes with non-zero fluxes. But this is a case of decay of the quantum system (contrary to elastic scattering in nuclear theory) and a similar process is used in the quantum description of the nucleation of the Universe and its evolution at its first stage.
The third correction of the semiclassical approximation (applied in a more restricted spatial region) transforms a real wave function into a complex one (see, for example, (46,11) in~\cite{Landau.v3.1989}) and now we see that such a picture without propagation of the wave under the barrier is only approximation.
So, the first conclusion is: the semiclassical approaches are less effective for a (detailed) study of tunneling processes with \underline{strong} propagation of waves (i.~e. for such spatial regions under barrier where the phenomenon of tunneling is present ``strongly'').

\item
The semiclassical wave function in the above barrier region is complex. From this we obtain \emph{non-conservation of the flux} in the transition through the boundary at the turning point.
Here, the flux calculated in the second semiclassical approximation describes propagation of the Universe in a ``classical sense'' but starting from the turning point, while exact calculations give conservation of the flux in the whole spatial region starting from zero.

\item
The flux non-conservation breaks down the correct change of the phase of wave function along the spatial axes (one may suppose that this violation can be essential). So, we come to the next characteristic --- the \emph{phase of the wave function} --- which can be less correctly estimated in the semiclassical approximation.

\item
From this we come to the \emph{phase time}, which is defined on the basis of the phase of the wave function and which is used in quantum dynamical theories. Thus, we come to the supposition that the semiclassical approach is less effective in a study of quantum dynamics (especially in the tunneling region).

\item
The phase includes information about \emph{interaction between barrier and wave}, and it can be interesting in a detailed study.

\item
\emph{The boundary condition} is imposed in the non-semiclassical region while we construct the outgoing wave in the semiclassical limit. One can suppose that the connection between the boundary condition and the formulation of the wave (for the same exact potential form) should be realized more accurately in the direct quantum approach.

\end{itemize}
Therefore, in order to estimate the probability of the nucleation of the Universe in the framework of the quantum approach with the highest accuracy, the necessity of the construction of a direct (non-semiclassical) method for an unambiguous construction of the wave in the region close to the escape point has become clear. A possible answer to the question above may be \emph{by the semiclassical approach some characteristics are determined very well, other ones can be determined less accurately and correctly}.

In this paper new strict and weak definitions of the wave propagating inside a potential of arbitrary shape and construction on this basis of the total wave function for cosmological potentials are proposed. After a short description of a simple variant of the quantum model of the evolution of the Universe in Sec.~\ref{sec.introduction}, a new method of the direct (non-semiclassical) determination of the stationary wave function at zero energy is presented and its behavior in the barrier region, near the escape point and in the asymptotic region, is analyzed in Sec.~\ref{sec.3}, and the barrier penetrability by the proposed approach is estimated with comparison with semiclassical calculations in Sec.~\ref{sec.4}.


\section{Cosmological model in the Friedmann--Robertson--Walker metric
\label{sec.introduction}}

One can describe the nucleation of the Universe by a simple model. Here, we shall consider the simplest variant of the homogeneous and isotropic Universe.
We consider the \emph{Friedmann--Robertson--Walker (FRW) metric} which defines the most general form of 4-dimensional spacetime with spherical spatial symmetry (see Ref.~\cite{Weinberg.1975}, p.~438; also see
Refs.~\cite{Rubakov.RTN2005,Linde.2005,Trodden.TASI-2003,Brandenberger.1999}):
%
\begin{equation}
\begin{array}{cccccc}
  ds^{2} = - dt^{2} + a^{2}(t) \cdot \biggl(\displaystyle\frac{dr^{2}}{h(r)} + r^{2} (d\theta^{2} + \sin^{2}{\theta} \, d\phi^{2}) \biggr), &
  h(r) = 1-kr^{2},
\end{array}
\label{eq.intro.1}
\end{equation}
where $t$ and $r$, $\theta$, $\phi$ are time and space spherical coordinates,
the signature of the metric is $(-,+,+,+)$ as in Ref.~\cite{Trodden.TASI-2003} (see~p.~4),
$a(t)$ is an unknown function of time and $k$ is a constant, the value of which equals $+1$, $0$ or $-1$, with appropriate choice of units for $r$. Further, we shall use the following system of units: $\hbar = c = 1$.
For $k = -1$, 0 the space is infinite (Universe of open type), and for $k=+1$ the space is finite (the Universe of closed type).
For $k=1$ one can describe the space as a sphere with radius $a(t)$ embedded in a 4-dimensional Euclidian space, and the function $a(t)$ is referred to as the \emph{``radius of the Universe''}. In the general case, it is called the \emph{cosmic scale factor}. This function contains information of the dynamics of the expansion of the Universe, and therefore its determination is an actual task.

One can find the function $a(t)$ using the Einstein equations in this metric (we use the signs according to the chosen signature, as in Ref.~\cite{Trodden.TASI-2003} p.~8; the Greek symbols $\mu$ and $\nu$ denote any of the four coordinates $t$, $r$, $\theta$ and $\phi$):
\begin{equation}
  R_{\mu\nu} - \displaystyle\frac{1}{2} \, g_{\mu\nu} \, R = 8\pi \: G\, T_{\mu\nu},
\label{eq.intro.2}
\end{equation}
where $R_{\mu\nu}$ is the Ricci tensor, $R$ is the scalar curvature, $T_{\mu\nu}$ is the energy-momentum tensor, and $G$ is Newton's constant.

Substituting the components of the Ricci tensor $R_{\mu\nu}$, the scalar curvature $R$, and the components of the energy-momentum tensor $T_{\mu\nu}$ (see Ref.~\cite{Trodden.TASI-2003}, p.~8)
%
%
%
%
into the equation (\ref{eq.intro.2}) at $\mu=\nu=0$, we obtain the \emph{Friedmann equation} (see p.~8 in Ref.~\cite{Trodden.TASI-2003}; p.~3 in Ref.~\cite{Brandenberger.1999}; p.~2 in Ref.~\cite{Vilenkin.1995}):
\begin{equation}
  \dot{a}^{2} + k - H^{2} a^{2} = 0,
\label{eq.intro.15}
\end{equation}
where
\begin{equation}
  H = \sqrt{\displaystyle\frac{8\pi G}{3} \cdot \rho},
\label{eq.intro.16}
\end{equation}
and $\rho$ is the energy density. Equation (\ref{eq.intro.15}) determines the function $a(t)$, describing the classical dynamics of the extension of the Universe, as determined by the energy density $\rho$.

Let us define the \emph{action} for the model. Usually, in the construction of the cosmological model the action must include both the geometry (curvature) of spacetime and matter fields. There are various papers proposing different variants of the cosmological models with inclusion of different types of the matter fields. In the given paper, in the construction of the action we restrict ourselves to the simplest component of the matter fields --- the vacuum only, directing the main attention to the development of the method for calculation of the wave function. We define the action as in Ref.~\cite{Vilenkin.1995} (see~(1), p.~2):
\begin{equation}
\begin{array}{ll}
  S = \displaystyle\int \sqrt{-g} \: \biggl( \displaystyle\frac{R}{16 \pi G} - \rho \biggr) \; dx^{4}, &

  R = \displaystyle\frac{6\dot{a}^{2} + 6a\ddot{a} + 6k}{a^{2}}.
\end{array}
\label{eq.intro.17}
\end{equation}
%
Substituting the scalar curvature, rewriting the energy density $\rho$ through $H$ according to (\ref{eq.intro.16}), and integrating the $\ddot{a}$ term in the resulting expression by parts with respect to $t$, we obtain the \emph{lagrangian} (see Ref.~\cite{Vilenkin.1995}, (11), p.~4):
\begin{equation}
  L(a, \dot{a}) = \displaystyle\frac{a}{2} \Bigl(k -  \dot{a}^{2} - H^{2} a^{2} \Bigr).
\label{eq.intro.18}
\end{equation}
Here, we shall consider the variables $a$ and $\dot{a}$ as generalized coordinate and velocity, respectively. Let us find the generalized momentum conjugate to $a$:
\begin{equation}
  p_{a} = \displaystyle\frac{\partial L(a, \dot{a})}{\partial \dot{a}} = -a\dot{a}.
\label{eq.intro.19}
\end{equation}
Further, we shall study the Universe of closed type only ($k=+1$). We then obtain the \emph{hamiltonian}:
\begin{equation}
  h(a, p_{a}) = p\,\dot{a} - L(a,\dot{a}) = - \displaystyle\frac{1}{2a} \Bigl( p_{a}^{2} + a^{2} - H^{2} a^{4} \Bigr).
\label{eq.intro.20}
\end{equation}
This hamiltonian describes the classical dynamics of the Universe, which serves as a basis for classical cosmological models.

The passage to the quantum description of the evolution of the Universe is obtained by the standard procedure of canonical quantization in the Dirac formalism for systems with constraints. We do the following.

\begin{itemize}
\item
We suppose that the variables $a$, $p_{a}$ and functions of them (for example, the hamiltonian) are operators, acting on a new function of state, named the \emph{wave function of the Universe}, $\varphi(a)$.

\item
We change the generalized coordinate $a$ and momentum $p_{a}$ into operators:
\begin{equation}
\begin{array}{cc}
  a \to \hat{a} = a, &
  p_{a} \to \hat{p_{a}} = -i \displaystyle\frac{\partial}{\partial a}.
\end{array}
\label{eq.intro.21}
\end{equation}
Here, the rule of the canonical correspondence between classical and quantum Poisson brackets is fulfilled.

\item
In order to take into account (\ref{eq.intro.15}) in the construction of the quantum model, we consider it as a constraint imposed on this quantum system, which gives us the eigenvalues (and eigenfunctions) of the operators which correspond most closely to fulfillment of this equation in the quantum case.

\end{itemize}

\begin{figure}[h]
\centering\includegraphics[width=7cm]{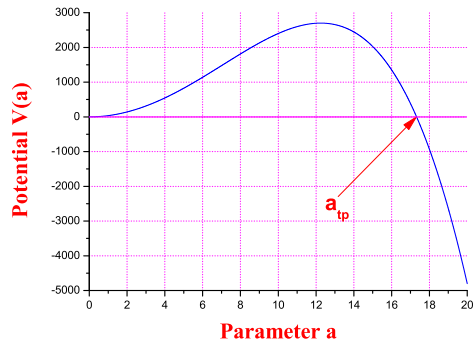}
\caption{The potential $V(a)$ with parameters $A=36$, $B=12\cdot \Lambda$ and $\Lambda=0.01$
(point $a_{tp} = 17.320508$).
\label{fig.1}}
\end{figure}

As a result of the quantization, we obtain the \emph{Wheeler--De Witt (WDW) equation} (see Ref.~\cite{Vilenkin.1995}, (16)--(17), in p.~4, \cite{Wheeler.1968,DeWitt.1967,Rubakov.2002.PRD}):
\begin{equation}
  \biggl( -\displaystyle\frac{\partial^{2}}{\partial a^{2}} + a^{2} - H^{2} a^{4} \biggr) \: \varphi(a) = 0.
\label{eq.intro.22}
\end{equation}
One can see that this equation looks similar to the one-dimensional stationary Schr\"{o}dinger equation on a semiaxis (of the variable $a$) at zero energy with potential
\begin{equation}
  V (a) = a^{2} - H^{2} a^{4}.
\label{eq.intro.23}
\end{equation}
Taking into account that in different papers different coefficients of $a^{2}$ and $a^{4}$ are used in the potential (\ref{eq.intro.23}) (as a result of a different construction of the action), for convenience, we rewrite this potential in a generalized form:
\begin{equation}
  V (a) = A\, a^{2} - B\, a^{4}.
\label{eq.intro.24}
\end{equation}
So, in particular, for the choice of parameters in Ref.~\cite{AcacioDeBarros.2006}
\begin{equation}
\begin{array}{ccc}
  A = 36, & B = 12 \cdot \Lambda, & \Lambda = 0.01,
\end{array}
\label{eq.intro.25}
\end{equation}
the potential is shown in Fig.~\ref{fig.1}.
Let us denote the point of intersection between the potential $V(a)$ and the $a$ axis as $a_{tp}$. Since we shall be studying the process of the escape from the barrier at zero energy, let us call the point $a_{tp}$ \emph{the escape point}.

After the passage from the classical description of the Universe evolution to the quantum one, all the information of the dynamics of the Universe expansion in the early stage and its further evolution, which was previously contained in the lagrangian (or hamiltonian), passes into the wave function $\varphi(a)$. Therefore, many papers today are devoted to the study of this function and any approach to its determination with the highest accuracy may be of interest.


\section{Determination of the wave function
\label{sec.3}}

Today, the two most prevailing ways for the determination of the wave function $\varphi(a)$ are the Feynman formalism of path integrals in multidimensional spacetime, developed by the Cambridge group and other researchers, called the \emph{``Hartle--Hawking method''} \cite{Hartle.1983.PRD}), and a method based on direct consideration of tunneling in 4-dimensional Euclidian spacetime, called the \emph{``Vilenkin method''} \cite{Vilenkin.1982.PLB,Vilenkin.1984.PRD,Vilenkin.1986.PRD,Vilenkin.1994.PRD}). We shall use the second way, which looks more natural and clearer for us. Here, the prevailing approach for the description of tunneling consists in its semiclassical consideration (for example, see Ref.~\cite{Vilenkin.1994.PRD,Rubakov.1999,Kuzmichev.1999.YAFIA,Kuzmichev.2002.EPJC}).
An attractive side of this approach is its simplicity in the construction of decreasing and increasing partial solutions for the wave function in the tunneling region, the outgoing wave function in the external region, and the possibility to define and estimate in a simply enough way the penetrability of the barrier, which can be used for obtaining the duration of the nucleation of the Universe.
We shall be interested in another way of direct non-semiclassical determination of the wave function $\varphi(a)$, in order to describe most accurately the process of the escape from the barrier, which describes the formation of the Universe and its further expansion at the first stage.

\subsection{The form of the wave function close to the escape point
\label{sec.3.1}}

For finding the wave function $\varphi(a)$ close to the escape point $a_{tp}$, we expand the potential $V(a)$ in (\ref{eq.intro.24}) in powers $q=a-a_{tp}$ (for small $q$) and restrict ourselves to the linear term only:
\begin{equation}
  V(q) = V_{0} + V_{1}q.
\label{eq.3.1.1}
\end{equation}
From (\ref{eq.intro.24}) we have
\begin{equation}
  a_{tp} = \sqrt{\displaystyle\frac{A}{B}},
\label{eq.3.1.2}
\end{equation}
\begin{equation}
\begin{array}{ccl}
  V_{0} & = & V(a=a_{tp}) = A a_{tp}^{2} - B a_{tp}^{4} = 0, \\
  V_{1} & = &
  2a_{tp} \: (A - 2B a_{tp}^{2}) =
  2 \sqrt{\displaystyle\frac{A}{B}} \cdot (A - 2A) =
  - 2A \sqrt{\displaystyle\frac{A}{B}},
\end{array}
\label{eq.3.1.3}
\end{equation}
and the Schr\"{o}dinger equation with the potential $V(q)$ with the new variable $q$ becomes
\begin{equation}
  -\displaystyle\frac{d^{2}}{dq^{2}}\, \varphi(q) - |V_{1}|q \: \varphi(q) = 0.
\label{eq.3.1.4}
\end{equation}
Changing the variable to
\begin{equation}
  \xi = |V_{1}|^{1/3} q,
\label{eq.3.1.5}
\end{equation}
(\ref{eq.3.1.4}) is transformed into
\begin{equation}
  \displaystyle\frac{d^{2}}{d\xi^{2}}\, \varphi(\xi) + \xi \: \varphi(\xi) = 0.
\label{eq.3.1.6}
\end{equation}

From quantum mechanics we know two linearly independent exact solutions for the function $\varphi(\xi)$ in this equation --- these are the \emph{Airy functions} ${\rm Ai}\,(\xi)$ and ${\rm Bi}\,(\xi)$ (for example, see Ref.~\cite{Abramowitz.1964}, p.~264--272, 291--294).
Expansions of these functions into power series at small $\xi$, their asymptotic expansions at large $|\xi|$, their representations through Bessel functions, zeroes and their asymptotic expansions are known. We have some integrals of these functions, and also the form of the Airy functions in the semiclassical approximation (which can be applied at large $|\xi|$).
In some problems of the analysis of finite solutions $\varphi(\xi)$ in the whole range of $\xi$ it is convenient to use the integral representations of the Airy functions (see (10.4.32) in Ref.~\cite{Abramowitz.1964}, p.~265, $a=1/3$, taking into account (10.4.1)):
\begin{equation}
\begin{array}{ccl}
  {\rm Ai} \, (\pm\xi) & = &
  \displaystyle\frac{1}{\pi} \displaystyle\int\limits_{0}^{+\infty}
  \cos{\biggl(\displaystyle\frac{u^{3}}{3} \mp \xi u \biggr)} \; du, \\

  {\rm Bi}\, (\pm\xi) & = &
  \displaystyle\frac{1}{\pi} \displaystyle\int\limits_{0}^{+\infty}
  \biggl[
    \exp{\biggl(-\displaystyle\frac{u^{3}}{3} \mp\xi u \biggr)} +
    \sin{\biggl(\displaystyle\frac{u^{3}}{3} \mp\xi u \biggr)}
  \biggr] \; du.
\end{array}
\label{eq.3.1.7}
\end{equation}

Furthermore, we shall be interested in the solution of the function $\varphi(\xi)$ which most accurately describes the \emph{outgoing wave} in the range of $a$ close to the point $a_{tp}$. However, it is not clear what the wave looks like near the point $a_{tp}$ in the potential studied in general, and which linear combination of the functions ${\rm Ai}\,(\xi)$ and ${\rm Bi}\,(\xi)$ defines it in the most accurate way.

The clearest and most natural understanding of the outgoing wave is given by the semiclassical consideration of the tunneling process. However, at the given potential the semiclassical approach allows us to define the outgoing wave in the asymptotic region only (while we can join solutions in the proximity of $a_{tp}$ by Airy functions). But it is not clear whether the wave, defined in the asymptotic region, remains outgoing near the $a_{tp}$. During the whole path of its propagation outside the barrier the wave interacts with the potential, which must inevitably lead to a deformation of its shape (like to appearance of a phase shift in the scattering of a wave by a radial potential caused by interaction in scattering theory). Could it turn out that the potentials used in cosmological models give a \underline{significantly larger change of the shape of the wave caused by interaction} in a comparison with the potentials used, for example, for the description of nuclear collisions in the framework of scattering theory? Moreover, for the given potential there is a problem with obtaining convergence in the calculation of the partial solutions for the wave function in the asymptotic region. According to our calculations, a small change in the range of the definition of the wave in the asymptotic region leads to a significant increase of errors, which requires one to increase the accuracy of the calculations.
Therefore, we shall be looking for a way of defining the outgoing wave not in the asymptotic region, but in the closest vicinity of the point of escape, $a_{tp}$. In a search of solutions close to the point $a_{tp}$, i.~e. at small enough $|\xi|$, the validity of the semiclassical method breaks down as $|\xi|$ approaches zero. Therefore, we shall not use the semiclassical approach in this paper.

Assuming the potential $V(a)$ to have an arbitrary form, we define the wave at the point $a_{tp}$ in the following way.

\begin{definition}[strict definition of wave]
The wave is such a linear combination of two partial solutions of the wave function that the change of the modulus $\rho$ of this wave function is closest to constant under variation of $a$:
\begin{equation}
  \displaystyle\frac{d^{2}}{da^{2}} \rho(a) \biggl|_{a=a_{tp}} \to 0.
\label{eq.3.1.8}
\end{equation}

\end{definition}
According to this definition, the real and imaginary parts of the total wave function have the mutually closest behaviors under the same variation of $a$, and the difference between possible maximums and minimums of the modulus of the total wave function is the smallest. For some types of potentials (in particular, for a rectangular barrier) it is more convenient to define the wave less strongly.
\begin{definition}[weak definition of wave]
The wave is such a linear combination of two partial solutions of wave function that the modulus $\rho$ changes minimally under variation of $a$:
\begin{equation}
  \displaystyle\frac{d}{da} \rho(a) \biggl|_{a=a_{tp}} \to 0.
\label{eq.3.1.9}
\end{equation}
\end{definition}
According to this definition, the change of the wave function caused by variation of $a$ is characterized mainly by its phase (which can characterize the interaction between the wave and the potential).

Subject to this requirement, we shall look for the solution for the function $\varphi(\xi)$ in the following form:
\begin{equation}
\begin{array}{ccl}
  \varphi (\xi) & = & N \cdot \Psi(\xi), \\
  \Psi (\xi) & = &
  \displaystyle\int\limits_{0}^{u_{\rm max}} \exp{i \Bigl(-\displaystyle\frac{u^{3}}{3} + f(\xi) u \Bigr)} \; du,
\end{array}
\label{eq.3.1.10}
\end{equation}
where $N$ is a normalization factor,
$f(\xi)$ is an unknown continuous function satisfying $f(\xi) \to {\rm const}$ as $\xi \to 0$,
and $u_{\rm max}$ is the unknown upper limit of integration. In such a solution, the real part of the function $f(\xi)$ gives a contribution to the phase of the integrand function only, while the imaginary part of $f(\xi)$ deforms its modulus.

Let us find the first and second derivatives of the function $\Psi(\xi)$ (a prime denotes a derivative with respect to $\xi$):
\begin{equation}
\begin{array}{ccl}
  \displaystyle\frac{d}{d\xi} \Psi (\xi) & = &
    i \displaystyle\int\limits_{0}^{u_{\rm max}} f^{\prime} u \;
    \exp{i \Bigl(-\displaystyle\frac{u^{3}}{3} + f(\xi) u \Bigr)} \; du, \\
  \displaystyle\frac{d^{2}}{d\xi^{2}} \Psi (\xi) & = &
    \displaystyle\int\limits_{0}^{u_{\rm max}} \Bigl(if^{\prime\prime} u - (f^{\prime})^{2} u^{2} \Bigr) \:
    \exp{i \Bigl(-\displaystyle\frac{u^{3}}{3} + f(\xi) u \Bigr)} \; du.
\end{array}
\label{eq.3.1.11}
\end{equation}
From this we obtain:
\begin{equation}
\begin{array}{c}
  \displaystyle\frac{d^{2}}{d\xi^{2}} \Psi (\xi) + \xi \Psi (\xi) =
  \displaystyle\int\limits_{0}^{u_{\rm max}} \Bigl(if^{\prime\prime} u - (f^{\prime})^{2} u^{2} + \xi \Bigr) \:
    \exp{i \Bigl(-\displaystyle\frac{u^{3}}{3} + f(\xi) u \Bigr)} \; du.
\end{array}
\label{eq.3.1.12}
\end{equation}

Considering solutions at small enough values of $|\xi|$, we represent $f(\xi)$ in the form of a power series:
\begin{equation}
  f(\xi) = \sum\limits_{n=0}^{+\infty} f_{n} \xi^{n},
\label{eq.3.1.13}
\end{equation}
where $f_{n}$ are constant coefficients. The first and second derivatives of $f(\xi)$ are
\begin{equation}
\begin{array}{l}
  f^{\prime}(\xi) = \displaystyle\frac{d}{d\xi} f(\xi) =
    \sum\limits_{n=1}^{+\infty} n f_{n} \; \xi^{n-1} = \sum\limits_{n=0}^{+\infty} (n+1) \: f_{n+1} \; \xi^{n}, \\
  f^{\prime\prime}(\xi) = \displaystyle\frac{d^{2}}{d\xi^{2}} f(\xi) =
    \sum\limits_{n=0}^{+\infty} (n+1)(n+2) \: f_{n+2} \; \xi^{n}.
\end{array}
\label{eq.3.1.14}
\end{equation}
Substituting these solutions into (\ref{eq.3.1.12}), we obtain
\begin{equation}
\begin{array}{c}
  \displaystyle\frac{d^{2}}{d\xi^{2}} \Psi (\xi) + \xi \Psi (\xi) =
  \displaystyle\int\limits_{0}^{u_{\rm max}}
    \Biggl\{
      \Bigl(2iu \: f_{2} - u^{2} \: f_{1}^{2} \Bigr) +
      \Bigl(6iu \: f_{3} - 4 u^{2} \: f_{1}f_{2} + 1 \Bigr) \: \xi + \\
    + \sum\limits_{n=2}^{+\infty} \Bigl[iu \: (n+1)(n+2) \: f_{n+2} -
      u^{2} \sum\limits_{m=0}^{n} (n-m+1)(m+1) \: f_{n-m+1}f_{m+1} \Bigr] \: \xi^{n}
    \Biggr\} \times \\
    \times \exp{i \Bigl(-\displaystyle\frac{u^{3}}{3} + fu \Bigr)} \; du = 0.
\end{array}
\label{eq.3.1.15}
\end{equation}

Considering this expression at small $|\xi|$, at the first step we use the following approximation:
\begin{equation}
\begin{array}{ccc}
  \exp{i \Bigl(-\displaystyle\frac{u^{3}}{3} + fu \Bigr)} & \to &
  \exp{i \Bigl(-\displaystyle\frac{u^{3}}{3} + f_{0}u \Bigr)}.
\end{array}
\label{eq.3.1.16}
\end{equation}
Requiring the condition (\ref{eq.3.1.15}) to be satisfied for different $\xi$ with different powers $n$, we obtain the following system:
\begin{equation}
\begin{array}{cl}
\xi^{0}: &
  \displaystyle\int\limits_{0}^{u_{\rm max}}
    \Bigl(2iu \: f_{2} - u^{2} \: f_{1}^{2} \Bigr) \:
    \exp{i \Bigl(-\displaystyle\frac{u^{3}}{3} + f_{0}u \Bigr)} \; du = 0, \\

\xi^{1}: &
  \displaystyle\int\limits_{0}^{u_{\rm max}}
      \Bigl(6iu \: f_{3} - 4 u^{2} \: f_{1}f_{2} + 1 \Bigr) \:
    \exp{i \Bigl(-\displaystyle\frac{u^{3}}{3} + f_{0}u \Bigr)} \; du = 0, \\

\xi^{n}: &
  \displaystyle\int\limits_{0}^{u_{\rm max}}
    \Bigl[iu \: (n+1)(n+2) \: f_{n+2} -
    u^{2} \sum\limits_{m=0}^{n} (n-m+1)(m+1) \: f_{n-m+1}f_{m+1} \Bigr] \times \\
   &  \times \exp{i \Bigl(-\displaystyle\frac{u^{3}}{3} + f_{0}u \Bigr)} \; du = 0.
\end{array}
\label{eq.3.1.17}
\end{equation}

Assuming the coefficients $f_{0}$ and $f_{1}$ to be given, we find the following solutions for the unknown $f_{2}$, $f_{3}$ and $f_{n}$:
\begin{equation}
\begin{array}{cc}
  f_{2} = \displaystyle\frac{f_{1}^{2}}{2i} \cdot \displaystyle\frac{J_{2}}{J_{1}}, &
  \hspace{8mm}
  f_{3} = \displaystyle\frac{4 f_{1}f_{2}\, J_{2} - J_{0}} {6i\, J_{1}},
\end{array}
\label{eq.3.1.18}
\end{equation}
\begin{equation}
  f_{n+2} = \displaystyle\frac{\sum\limits_{m=0}^{n} (n-m+1)(m+1) \: f_{n-m+1}f_{m+1}} {i \: (n+1)(n+2)} \cdot
    \displaystyle\frac{J_{2}} {J_{1}},
\label{eq.3.1.19}
\end{equation}
where the following notations for the integrals have been introduced:
\begin{equation}
\begin{array}{c}
  J_{0} =
    \displaystyle\int\limits_{0}^{u_{\rm max}}
    \exp{i \Bigl(-\displaystyle\frac{u^{3}}{3} + f_{0}u \Bigr)} \; du, \\
  J_{1} =
    \displaystyle\int\limits_{0}^{u_{\rm max}}
    u \: \exp{i \Bigl(-\displaystyle\frac{u^{3}}{3} + f_{0}u \Bigr)} \; du, \\
  J_{2} =
    \displaystyle\int\limits_{0}^{u_{\rm max}}
    u^{2} \: \exp{i \Bigl(-\displaystyle\frac{u^{3}}{3} + f_{0}u \Bigr)} \; du.
\end{array}
\label{eq.3.1.20}
\end{equation}

Thus, we see that the solution (\ref{eq.3.1.10}) for the function $\varphi(\xi)$ has arbitrariness in the choice of the unknown coefficients $f_{0}$, $f_{1}$ and the upper limit of integration $u_{\rm max}$. However, the solutions found, (\ref{eq.3.1.18}) and (\ref{eq.3.1.19}), define the function $f(\xi)$ so as to ensure that the equality (\ref{eq.3.1.6}) is \underline{exactly} satisfied in the region of $a$ close to the escape point $a_{tp}$.
This proves that \emph{the function $\varphi(\xi)$ in the form (\ref{eq.3.1.10}) with an arbitrary choice of $f_{0}$, $f_{1}$ and $u_{\rm max}$ is the solution of the Schr\"{o}dinger equation (\ref{eq.3.1.10}) near the escape point $a_{tp}$ and its accuracy is maximal at $a=a_{tp}$}. In order to bring the solution $\Psi(\xi)$ into the well-known form of the Airy functions, ${\rm Ai}\, (\xi)$ and ${\rm Bi}\, (\xi)$,
we select
\begin{equation}
\begin{array}{cc}
  f_{0} = 0, &
  f_{1} = 1.
\end{array}
\label{eq.3.1.21}
\end{equation}
At such choice of the coefficients $f_{0}$ and $f_{1}$, the integrand function in the solution (\ref{eq.3.1.10}) up to $\xi^{2}$ has a constant modulus and a varying phase (the coefficient $f_{2}$ deforms the modulus, but it is fulfilled at $\xi^{3}$). Therefore, one can expect that the solution (\ref{eq.3.1.10}) at the escape point $a_{tp}$ describes the wave with the proper shape.


\subsection{Calculation of two partial solutions of the wave function
\label{sec.3.2}}

Not knowing the behavior of the wave function and its derivative as a function of $a$, we shall be looking for their two partial solutions independently in the following way. At first, we shall define the wave function and its derivative at a selected point, and we shall calculate the wave function and its derivative in the region close enough to this point using the method of starting of the solution; this is presented in the next two sections. Here, for the partial solution, which increases in the barrier region, we use the starting point to be $a=0$, and for the second partial solution, which decreases in the barrier region, we select the starting point to be the escape point $a_{tp}$. Further, we shall calculate the wave function and its derivative in the whole required range of $a$ using the method of continuation of the solution, for which we select the Numerov method with a constant step.

\subsubsection{Determination of the wave function close to zero. A regular solution
\label{sec.3.2.1}}

We shall be looking for the regular solution for the wave function close to $a=0$.
Let's write the wave function close to $a=0$ in the following form:
\begin{equation}
  \varphi(a) = c_{1} \sum\limits_{n=0}^{+\infty} a_{n} \: a^{n},
\label{eq.3.2.1.1}
\end{equation}
where $a_{n}$ are constant coefficients.

One can find the unknown $a_{n}$ from the Schr\"{o}dinger equation with the potential (\ref{eq.intro.24}). From (\ref{eq.3.2.1.1}) we obtain the first and second derivatives of the wave function:
\begin{equation}
\begin{array}{ccl}
  \varphi^{\prime}(a) & = &
    c_{1} \sum\limits_{n=1}^{+\infty} n \, a_{n} \; a^{n-1} =
    c_{1} \sum\limits_{n=0}^{+\infty} (n+1) \, a_{n+1} \; a^{n}, \\
  \varphi^{\prime\prime}(a) & = &
    c_{1} \sum\limits_{n=1}^{+\infty} (n+1)\, n \, a_{n+1} \; a^{n-1} =
    c_{1} \sum\limits_{n=0}^{+\infty} (n+2)(n+1) \, a_{n+2} \; a^{n}.
\end{array}
\label{eq.3.2.1.2}
\end{equation}
The Schr\"{o}dinger equation and the potential have the forms
\begin{equation}
\begin{array}{cc}
  -\varphi^{\prime\prime}(a) + V(a) \: \varphi(a) = 0, &
  V(a) = A\, a^{2} - B \, a^{4}.
\end{array}
\label{eq.3.2.1.3}
\end{equation}
Substituting (\ref{eq.3.2.1.1}) and (\ref{eq.3.2.1.2}) for the wave function and its second derivative, we obtain
\[
\begin{array}{l}
  \varphi^{\prime\prime}(a) = c_{1} \sum\limits_{n=0}^{+\infty} (n+2)\,(n+1)\,a_{n+2} \; a^{n} = \\
  = \Bigl( A \,a^{2} - B\,a^{4} \Bigr) \: \varphi(a) =
  \Bigl( A \,a^{2} - B\,a^{4} \Bigr) \: c_{1} \sum\limits_{n=0}^{+\infty} a_{n}\, a^{n} = \\
  = c_{1} \biggl\{ A \sum\limits_{n=0}^{+\infty} a_{n} \,a^{n+2} - B \sum\limits_{n=0}^{+\infty} a_{n} \, a^{n+4} \biggr \} =
  c_{1} \biggl\{ A \sum\limits_{n=2}^{+\infty} a_{n-2} \, a^{n} - B \sum\limits_{n=4}^{+\infty} a_{n-4} \, a^{n} \biggr \} = \\
  = c_{1} \biggl\{ A \Bigl(a_{0}\,a^{2} + a_{1}\,a^{3} \Bigr) +
  \sum\limits_{n=4}^{+\infty} \Bigl(A\, a_{n-2} - B\, a_{n-4} \Bigr) \; a^{n} \biggr \}
\end{array}
\]
or
\begin{equation}
\begin{array}{l}
  \sum\limits_{n=0}^{+\infty} (n+2)\,(n+1)\, a_{n+2} \; a^{n} =
  A \, \Bigl(a_{0}\,a^{2} + a_{1}\,a^{3}\Bigr) +
  \sum\limits_{n=4}^{+\infty} \Bigl(A\, a_{n-2} - B\, a_{n-4} \Bigr) \; a^{n}.
\end{array}
\label{eq.3.2.1.4}
\end{equation}
Let us write the expressions at $a^{n}$ with the same powers $n$:
\begin{equation}
\begin{array}{ccl}
  n=0:     & \to & 2 \cdot 1 \cdot a_{2} \cdot a^{0} = 0; \\
  n=1:     & \to & 3 \cdot 2 \cdot a_{3} \cdot a^{1} = 0; \\
  n=2:     & \to & 4 \cdot 3 \cdot a_{4} \cdot a^{2} = A \cdot a_{0} \cdot a^{2}; \\
  n=3:     & \to & 5 \cdot 4 \cdot a_{5} \cdot a^{3} = A \cdot a_{1} \cdot a^{3}; \\
  n \ge 4: & \to & (n+2)\,(n+1) \cdot a_{n+2} \cdot a^{n} = \Bigl( A\,a_{n-2} - B\,a_{n-4} \Bigr)\: a^{n}.
\end{array}
\label{eq.3.2.1.5}
\end{equation}
Thus we obtain recurrent relations for the calculation of the unknown $a_{n}$:
\begin{equation}
\begin{array}{ccccrl}
  a_{2} = 0, &
  a_{3} = 0, &
  a_{4} = \displaystyle\frac{A\, a_{0}}{12}, &
  a_{5} = \displaystyle\frac{A\, a_{1}}{20}, &
  a_{n+2} = \displaystyle\frac{A\,a_{n-2} - B\,a_{n-4}}{(n+1)\,(n+2)} & \mbox{at } n \ge 4.
\end{array}
\label{eq.3.2.1.6}
\end{equation}
Given values for $a_{0}$ and $a_{1}$, using (\ref{eq.3.2.1.6}) one can calculate all $a_{n}$ needed.

Analyzing (\ref{eq.3.2.1.1}) at zero, we find (here, we use $c_{1}=1$):
\begin{equation}
\begin{array}{cc}
  a_{0} = \varphi(0), & a_{1} = \varphi^{\prime}(0).
\end{array}
\label{eq.3.2.1.7}
\end{equation}
So the coefficients $a_{0}$ and $a_{1}$ determine the wave function and its derivative at zero. Let us write all possible cases:
\begin{equation}
\begin{array}{lllll}
  \varphi(0) = 0,   & \varphi^{\prime}(0) \ne 0 & \to & a_{0} = 0,   & a_{1} \ne 0, \\
  \varphi(0) \ne 0, & \varphi^{\prime}(0) = 0   & \to & a_{0} \ne 0, & a_{1} = 0, \\
  \varphi(0) \ne 0, & \varphi^{\prime}(0) \ne 0 & \to & a_{0} \ne 0, & a_{1} \ne 0.
\end{array}
\label{eq.3.2.1.8}
\end{equation}
One can consider expressions (\ref{eq.3.2.1.8}) as different boundary conditions for the wave function and its derivative at zero. Implying two different boundary conditions through $a_{0}$ and $a_{1}$ (in such a way that they locate the first node for the wave function at different places), we shall obtain two \emph{linearly independent partial solutions $\varphi_{1}(a)$ and $\varphi_{2}(a)$ for the wave function} close to zero. At the present stage, not knowing the coordinates of the maxima and coordinates of the next nodes for the partial solutions for the wave function, we select
\begin{equation}
\begin{array}{lllll}
  \varphi_{1}(0) = 0,   & \varphi_{1}^{\prime}(0) \ne 0 & \to & a_{0} = 0, & a_{1} =1, \\
  \varphi_{2}(0) \ne 0, & \varphi_{2}^{\prime}(0) = 0   & \to & a_{0} = 1, & a_{1} = 0.
\end{array}
\label{eq.3.2.1.9}
\end{equation}
However, analysis has shown that the error in the calculation of the decreasing component of the wave function can grow significantly when $a$ increases (depending on the choice of the parameters of the potential), which makes the calculation of that component inefficient. Therefore, in this way we shall obtain one partial solution for the wave function only, which increases in the barrier region (we select the first condition from (\ref{eq.3.2.1.9})).


\subsubsection{Determination of the wave function near an arbitrary selected point
\label{sec.3.2.2}}

Now we shall find a regular solution for the wave function near a point $a_{x}$, which can be selected arbitrarily on the semiaxis. We write such a wave function in the following form:
\begin{equation}
  \varphi(a) = c_{2} \sum\limits_{n=0}^{+\infty} b_{n} \: (a-a_{x})^{n}
  = c_{2} \sum\limits_{n=0}^{+\infty} b_{n} \: \bar{a}^{n},
\label{eq.3.2.2.1}
\end{equation}
where $b_{n}$ are new constant coefficients, and we introduced the new variable
\begin{equation}
  \bar{a} = a-a_{x}.
\label{eq.3.2.2.2}
\end{equation}
Let us find the first and second derivatives of the wave function (\ref{eq.3.2.2.1}):
\begin{equation}
\begin{array}{ccl}
  \varphi^{\prime}(a) & = &
    c_{2} \sum\limits_{n=1}^{+\infty} n \, b_{n} \; \bar{a}^{n-1} =
    c_{2} \sum\limits_{n=0}^{+\infty} (n+1) \, b_{n+1} \; \bar{a}^{n}, \\
  \varphi^{\prime\prime}(a) & = &
    c_{2} \sum\limits_{n=1}^{+\infty} (n+1)\, n \, b_{n+1} \; \bar{a}^{n-1} =
    c_{2} \sum\limits_{n=0}^{+\infty} (n+2)(n+1) \, b_{n+2} \; \bar{a}^{n}.
\end{array}
\label{eq.3.2.2.3}
\end{equation}

We rewrite the potential through the variable $\bar{a}$:
\begin{equation}
\begin{array}{ccl}
  V(a) & = & A\, a^{2} - B\, a^{4} = a^{2} (A - B\, a^{2}) = (\bar{a} + a_{x})^{2} (A - B\, (\bar{a} + a_{x})^{2}) =\\
       & = & (a_{x}^{2} + 2 a_{x}\bar{a} + \bar{a}^{2}) (A - B\, a_{x}^{2} - 2 B\, a_{x}\bar{a} - B\,\bar{a}^{2}) =\\
       & = & C_{0} + C_{1}\,\bar{a} + C_{2}\,\bar{a}^{2} + C_{3}\,\bar{a}^{3} + C_{4}\,\bar{a}^{4},
\end{array}
\label{eq.3.2.2.4}
\end{equation}
where
\begin{equation}
\begin{array}{ccl}
  C_{0} & = & A\, a_{x}^{2} - B \,a_{x}^{4}, \\
  C_{1} & = & 2a_{x}(A-B\,a_{x}^{2}) - 2B\,a_{x}^{3} = 2A\,a_{x} - 4B\,a_{x}^{3}, \\
  C_{2} & = & A - B\,a_{x}^{2} - 4B\, a_{x}^{2} - B\,a_{x}^{2} = A - 6B\,a_{x}^{2}, \\
  C_{3} & = & -2B\,a_{x} - 2B\,a_{x} = -4B\,a_{x}, \\
  C_{4} & = & - B.
\end{array}
\label{eq.3.2.2.5}
\end{equation}

Substituting the wave function (\ref{eq.3.2.2.1}), the second derivative of the wave function (\ref{eq.3.2.2.3}) and the potential (\ref{eq.3.2.2.4}) into the Schr\"{o}dinger equation, we obtain
\[
\begin{array}{l}
  \varphi^{\prime\prime}(a) = c_{2} \sum\limits_{n=0}^{+\infty} (n+2)\,(n+1)\,b_{n+2} \; \bar{a}^{n} = \\
  = \Bigl( A \,a^{2} - B\,a^{4} \Bigr) \: \varphi(a) =
  \Bigl( C_{0} + C_{1}\,\bar{a} + C_{2}\,\bar{a}^{2} + C_{3}\,\bar{a}^{3} + C_{4}\,\bar{a}^{4} \Bigr) \:
  c_{2} \sum\limits_{n=0}^{+\infty} b_{n}\, \bar{a}^{n} = \\

  = c_{2} \biggl\{ C_{0} \sum\limits_{n=0}^{+\infty} b_{n} \,\bar{a}^{n} +
    C_{1} \sum\limits_{n=0}^{+\infty} b_{n} \,\bar{a}^{n+1} +
    C_{2} \sum\limits_{n=0}^{+\infty} b_{n} \,\bar{a}^{n+2} +
    C_{3} \sum\limits_{n=0}^{+\infty} b_{n} \,\bar{a}^{n+3} +
    C_{4} \sum\limits_{n=0}^{+\infty} b_{n} \,\bar{a}^{n+4} \biggr \} = \\

  = c_{2} \biggl\{ C_{0} \sum\limits_{n=0}^{+\infty} b_{n} \,\bar{a}^{n} +
    C_{1} \sum\limits_{n=1}^{+\infty} b_{n-1} \,\bar{a}^{n} +
    C_{2} \sum\limits_{n=2}^{+\infty} b_{n-2} \,\bar{a}^{n} +
    C_{3} \sum\limits_{n=3}^{+\infty} b_{n-3} \,\bar{a}^{n} +
    C_{4} \sum\limits_{n=4}^{+\infty} b_{n-4} \,\bar{a}^{n} \biggr \} = \\

  = c_{2} \biggl\{ C_{0}\,b_{0} +
    \bigl( C_{0}\,b_{1} + C_{1}\,b_{0} \bigr) \;\bar{a} +
    \bigl( C_{0}\,b_{2} + C_{1}\,b_{1} + C_{2}\,b_{0} \bigr) \;\bar{a}^{2} +
    \bigl( C_{0}\,b_{3} + C_{1}\,b_{2} + C_{2}\,b_{1} + C_{3}\,b_{0} \bigr) \;\bar{a}^{3} + \\

    \; + \sum\limits_{n=4}^{+\infty}
      \bigl( C_{0}\,b_{n}+ C_{1}\,b_{n-1}+ C_{2}\,b_{n-2}+ C_{3}\,b_{n-3}+ C_{4}\,b_{n-4} \bigr)
      \;\bar{a}^{n} \biggr \}
\end{array}
\]
or
\begin{equation}
\begin{array}{l}
  \sum\limits_{n=0}^{+\infty} (n+2)\,(n+1)\,b_{n+2} \; \bar{a}^{n} = \\

  = C_{0}\,b_{0} +
    \bigl( C_{0}\,b_{1} + C_{1}\,b_{0} \bigr) \;\bar{a} +
    \bigl( C_{0}\,b_{2} + C_{1}\,b_{1} + C_{2}\,b_{0} \bigr) \;\bar{a}^{2} +
    \bigl( C_{0}\,b_{3} + C_{1}\,b_{2} + C_{2}\,b_{1} + C_{3}\,b_{0} \bigr) \;\bar{a}^{3} + \\
    \; + \sum\limits_{n=4}^{+\infty}
      \bigl( C_{0}\,b_{n}+ C_{1}\,b_{n-1}+ C_{2}\,b_{n-2}+ C_{3}\,b_{n-3}+ C_{4}\,b_{n-4} \bigr)
      \;\bar{a}^{n}.
\end{array}
\label{eq.3.2.2.6}
\end{equation}
We write the expressions at $\bar{a}^{n}$ with the same powers $n$:
\begin{equation}
\begin{array}{ccl}
  n=0:     & \to & 2 \cdot 1 \cdot b_{2} \cdot \bar{a}^{0} = C_{0} b_{0} \cdot \bar{a}^{0}; \\
  n=1:     & \to & 3 \cdot 2 \cdot b_{3} \cdot \bar{a}^{1} =
                     \bigl( C_{0}\,b_{1} + C_{1}\,b_{0} \bigr) \cdot \bar{a}^{1}; \\
  n=2:     & \to & 4 \cdot 3 \cdot b_{4} \cdot \bar{a}^{2} =
                     \bigl( C_{0}\,b_{2} + C_{1}\,b_{1} + C_{2}\,b_{0} \bigr) \cdot \bar{a}^{2}; \\
  n=3:     & \to & 5 \cdot 4 \cdot b_{5} \cdot \bar{a}^{3} =
                     \bigl( C_{0}\,b_{3} + C_{1}\,b_{2} + C_{2}\,b_{1} + C_{3}\,b_{0} \bigr) \cdot \bar{a}^{3}; \\
  n \ge 4: & \to & (n+2)\,(n+1) \cdot b_{n+2} \cdot \bar{a}^{n} = \
             \bigl( C_{0}\,b_{n}+ C_{1}\,b_{n-1}+ C_{2}\,b_{n-2}+ C_{3}\,b_{n-3}+ C_{4}\,b_{n-4} \bigr)
             \cdot \bar{a}^{n}.
\end{array}
\label{eq.3.2.2.7}
\end{equation}
As a result, we obtain recurrent relations for the determination of the unknown $b_{n}$:
\begin{equation}
\begin{array}{ccccrl}
  b_{2} = \displaystyle\frac{C_{0}\,b_{0}}{2}, &
  b_{3} = \displaystyle\frac{C_{0}\,b_{1} + C_{1}\,b_{0}}{6}, &
  b_{4} = \displaystyle\frac{C_{0}\,b_{2} + C_{1}\,b_{1} + C_{2}\,b_{0}}{12}, &
  b_{5} = \displaystyle\frac{C_{0}\,b_{3} + C_{1}\,b_{2} + C_{2}\,b_{1} + C_{3}\,b_{0}}{20}, &
\end{array}
\label{eq.3.2.2.8}
\end{equation}
\begin{equation}
\begin{array}{ccccrl}
  b_{n+2} =
  \displaystyle\frac{C_{0}\,b_{n}+ C_{1}\,b_{n-1}+ C_{2}\,b_{n-2}+ C_{3}\,b_{n-3}+ C_{4}\,b_{n-4}}
  {(n+1)\,(n+2)} & \mbox{with } n \ge 4.
\end{array}
\label{eq.3.2.2.9}
\end{equation}

From (\ref{eq.3.2.1.1}) and (\ref{eq.3.2.1.3}) we find (we use $c_{2}=1$)
\begin{equation}
\begin{array}{cc}
  b_{0} = \varphi(a_{x}), & b_{1} = \varphi^{\prime}(a_{x}).
\end{array}
\label{eq.3.2.2.10}
\end{equation}
Thus, using the coefficients $b_{0}$ and $b_{1}$, one can define the wave function and its derivative at the point $a_{x}$. Defining $b_{0}$ and $b_{1}$ and using (\ref{eq.3.2.2.9}), one can calculate all $b_{n}$ needed.

Analysis has shown that this way is more efficient (in comparison with the previous one) in the calculation of the decreasing component of the wave function. Therefore, using the escape point $a_{tp}$ as the starting point, we shall use this way to calculate the second partial solution, which is decreasing in the barrier region, and we select
\begin{equation}
\begin{array}{cc}
  b_{0} = 1, & b_{1} = -0.1.
\end{array}
\label{eq.3.2.2.11}
\end{equation}

\subsubsection{Determination of the first partial solution in the whole region, increasing in the region of tunneling
\label{sec.3.2.3}}

Picking values for the wave function and its derivative at $a=0$ and calculating them in the vicinity of that point, we shall calculate them in the whole range of $a$, using the Numerov method with a constant step. We see that such an approach allows one to obtain a convergent and stable partial solution, which is increasing in the tunneling region. In order to analyze to what extent this approach gives convergent (stable) solutions, for a comparison we use \cite{AcacioDeBarros.2006} with data presented for the modulus of the wave function (see (9) p.~5; we use the parameters of the potential in (\ref{eq.intro.25})).

\begin{figure}[h]
\centerline{
\includegraphics[width=75mm]{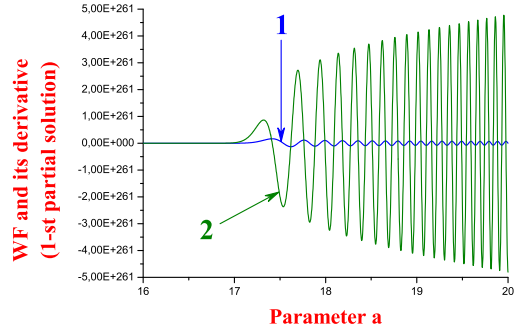}}
\caption{\small
The first partial solution for the wave function and its derivative (decreasing in the tunneling region):
curve 1, blue, is for the wave function;
curve 2, green, for the derivative of this wave function.
\label{fig.2}}
\end{figure}

The first partial solution for the wave function and its derivative, obtained by our calculations in the region of the escape point, is shown in Fig.~\ref{fig.2}. Comparing them with results in Ref.~\cite{AcacioDeBarros.2006}, we see that in our approach the wave function and its derivative look more continuous and have no divergences (this becomes more apparent after a detailed analysis of the selected regions).
In the figures one can see that starting from the point $a_{tp}$ the wave function decreases and its derivative increases with increasing $a$. One can see that the derivative looks significantly larger than the wave function.

\begin{figure}[h]
\centerline{
\includegraphics[width=75mm]{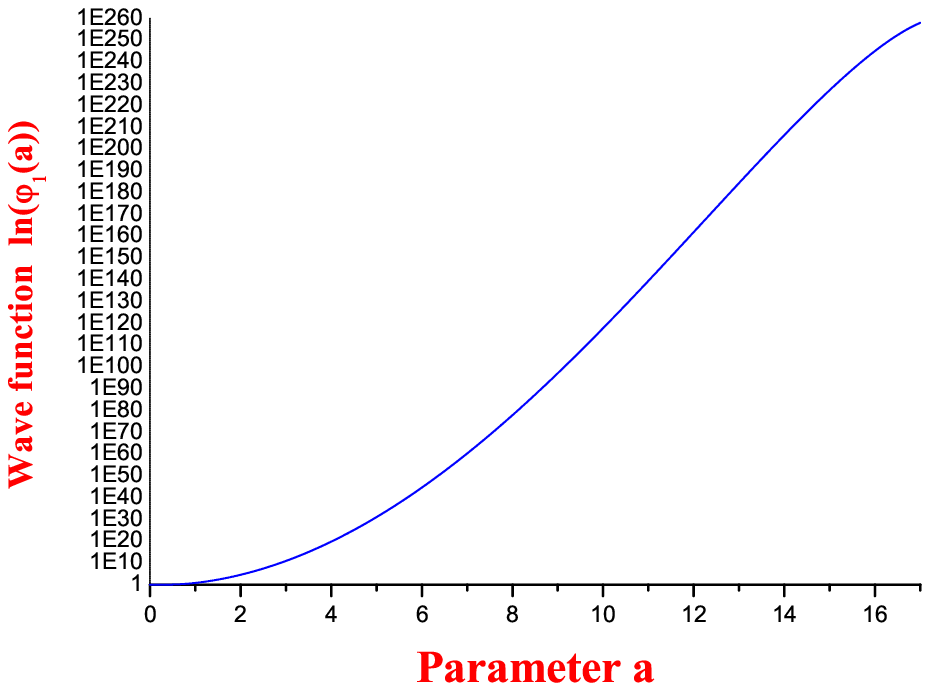}
\includegraphics[width=75mm]{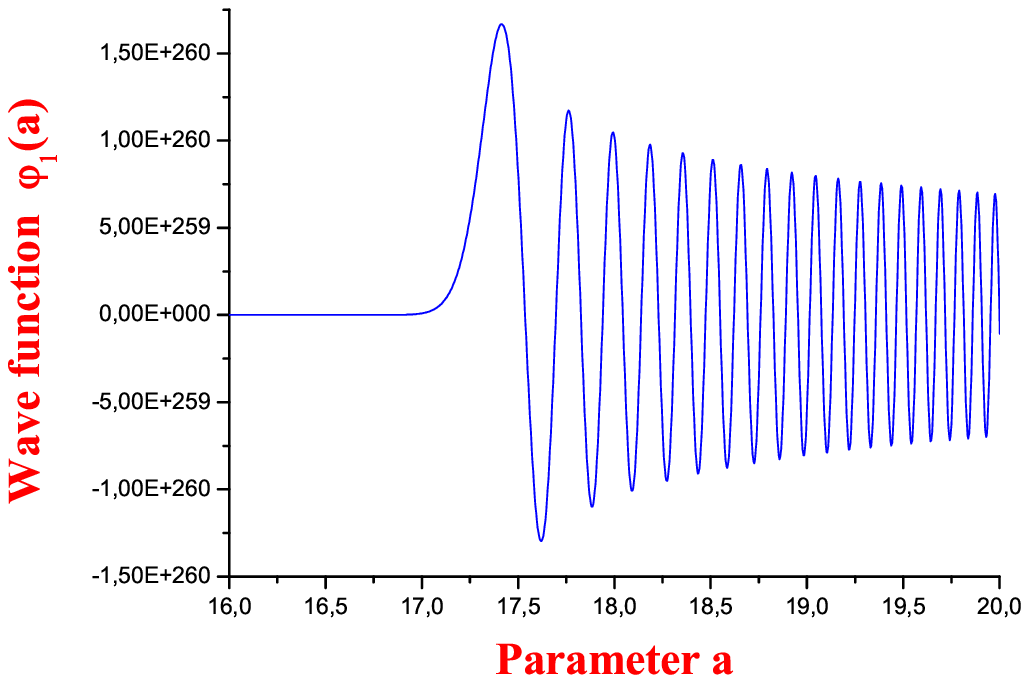}}
\caption{\small
The behavior of the wave function in the selected regions:
(a) is for the wave function in the tunneling region ($\ln{(\varphi_{1}(a))}$ is in vertical axis);
(b) for the wave function in the region of the escape point.
\label{fig.3}}
\end{figure}

Now let us consider the behavior of the wave function more carefully. In Fig.~\ref{fig.3} it is shown how it changes in the tunneling region and in the vicinity of the escape point. From the figures one can see that the wave function satisfies the rules of behavior of the wave functions inside sub-barrier and above the barrier region. Starting from very small $a$, the wave function increases monotonously (without any oscillations) with increasing $a$, which corresponds to the tunneling region (this becomes more obvious especially in a logarithmic presentation of the wave function; see the left figure). Further, oscillations have appeared in the wave function, which can be possible only inside the above barrier region (a smooth transition is shown in the right figure). The boundary of such a transition of the wave function must be the escape point $a_{tp}$ (our calculations give $a_{tp}=17.320508$,
which coincides with Ref.~\cite{AcacioDeBarros.2006}).

\begin{figure}[h]
\centerline{
\includegraphics[width=75mm]{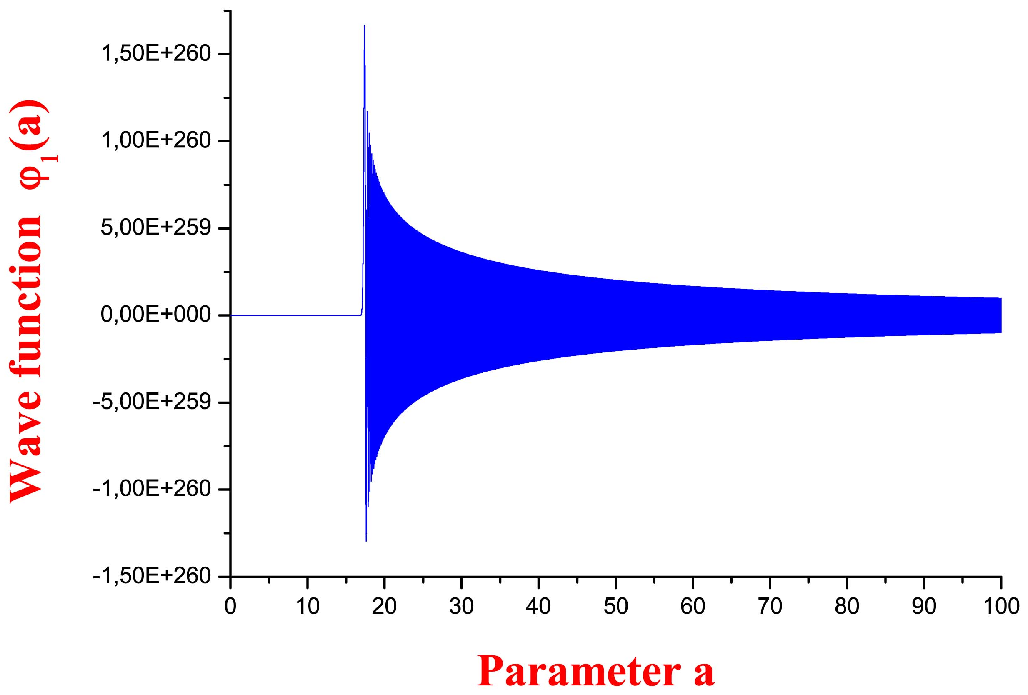}
\includegraphics[width=75mm]{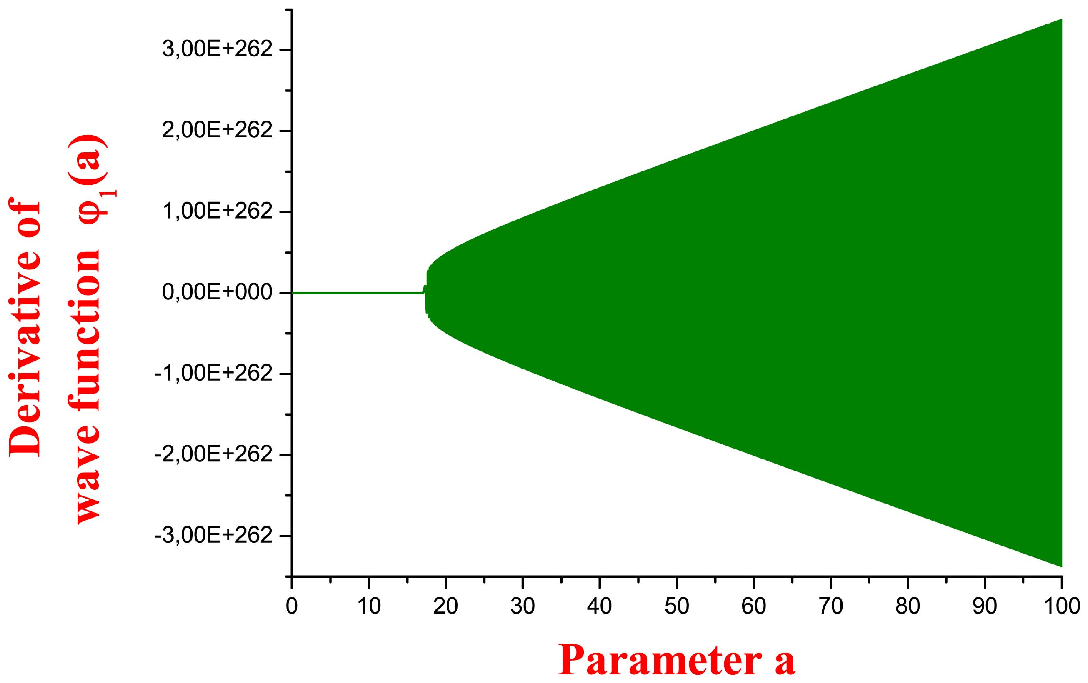}}
\caption{\small
Asymptotic behavior of the wave function (a) and its derivative (b).
\label{fig.4}}
\end{figure}

In Fig.~\ref{fig.4} it is shown how the wave function and its derivative vary at large values of $a$, i.~e. including a region which may be called an \emph{asymptotic} one. We obtain smooth continuous solutions running up to $a=100$ (in Ref.~\cite{AcacioDeBarros.2006} the maximum value is $a=30$).
An interesting peculiarity of the obtained solutions is an absolutely uniform straight increase of the maxima of the derivative of the wave function and a smooth decrease of the wave function with increasing $a$, which must point to the specific character of the expansion of the Universe in this quantum-mechanical approach.

Analyzing all the obtained figures, we find the following interesting property.

\noindent
\emph{With increasing $a$, the period of oscillations decreases uniformly both for the wave function and its derivative in the above barrier region!}

\begin{itemize}

\item
This explains why for sufficiently small increases of $a$ it becomes \underline{significantly more difficult} to calculate the convergent continuous solutions for the wave function and its derivative (where it is needed to significantly decrease the step)!

\item
Instead of the Numerov method with constant step, it can be more efficient to use another method of continuation of the solution, in which the step decreases uniformly with increasing of $a$.

\item
This information can be interesting for the choice or construction of new functions, by which the total wave function in the interesting region can be expanded with the highest efficiency (for example, this is why searching for solutions for the total wave function as an expansion in plane waves or spherical Bessel functions cannot give a sufficiently stable convergent result (with a small increase of the selected range of $a$)).
\end{itemize}

\subsubsection{Determination of the second partial solution on the whole region, decreasing in the tunneling region \label{sec.3.2.4}}

We find the second partial solution similarly to the first one, but with a different starting point. First, setting the value for the wave function and its derivative at the escape point $a_{tp}$, we find them in the vicinity of that point by the method from sec.~\ref{sec.3.2.2}. Further, using the Numerov method with a constant step, we find solutions in the whole region of $a$.

\begin{figure}[h]
\centerline{
\includegraphics[width=75mm]{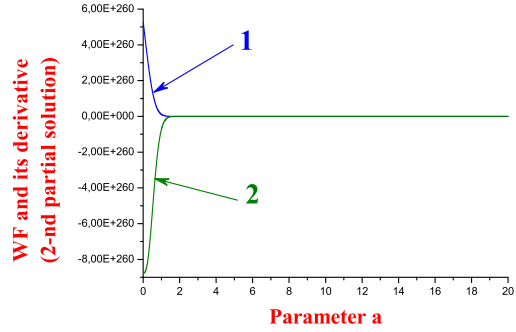}}
\caption{\small
The second partial solution for the wave function and its derivative, decreasing in the tunneling region:
curve 1, blue, is for the wave function;
curve 2, green, for the derivative of this wave function.
\label{fig.5}}
\end{figure}

In Fig.~\ref{fig.5} the wave function and its derivative in the region with the right boundary larger than point $a_{tp}$ are shown (parameters of the potential are from (\ref{eq.intro.25})). One can see from this figure
a strong decrease of the amplitudes of the wave function and its derivative with increasing $a$, which implies that the solutions obtained are \emph{decreasing in the tunneling region}.

\begin{figure}[h]
\centerline{
\includegraphics[width=55mm]{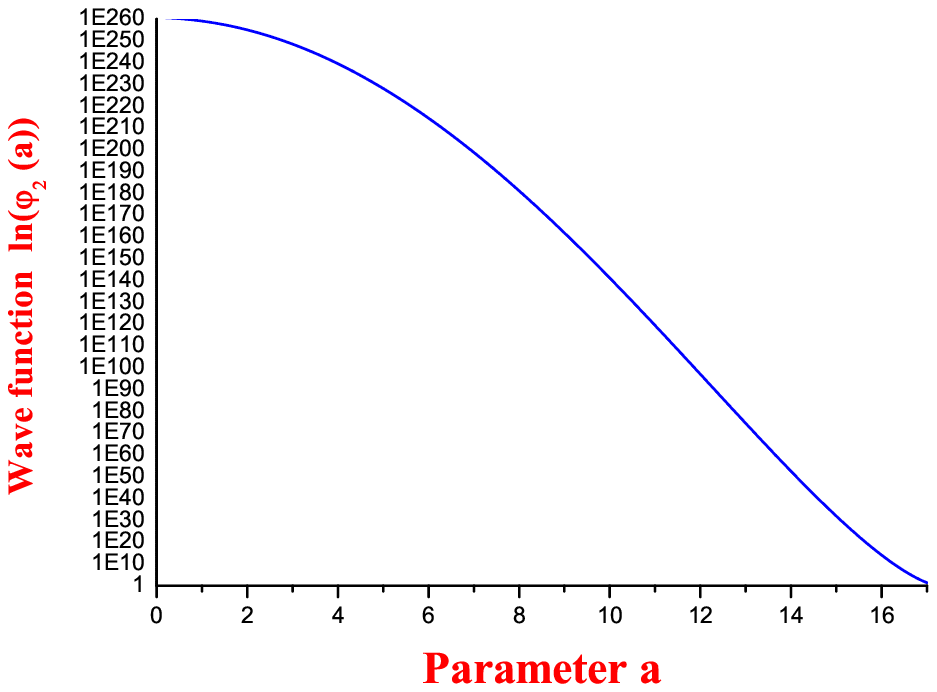}
\includegraphics[width=55mm]{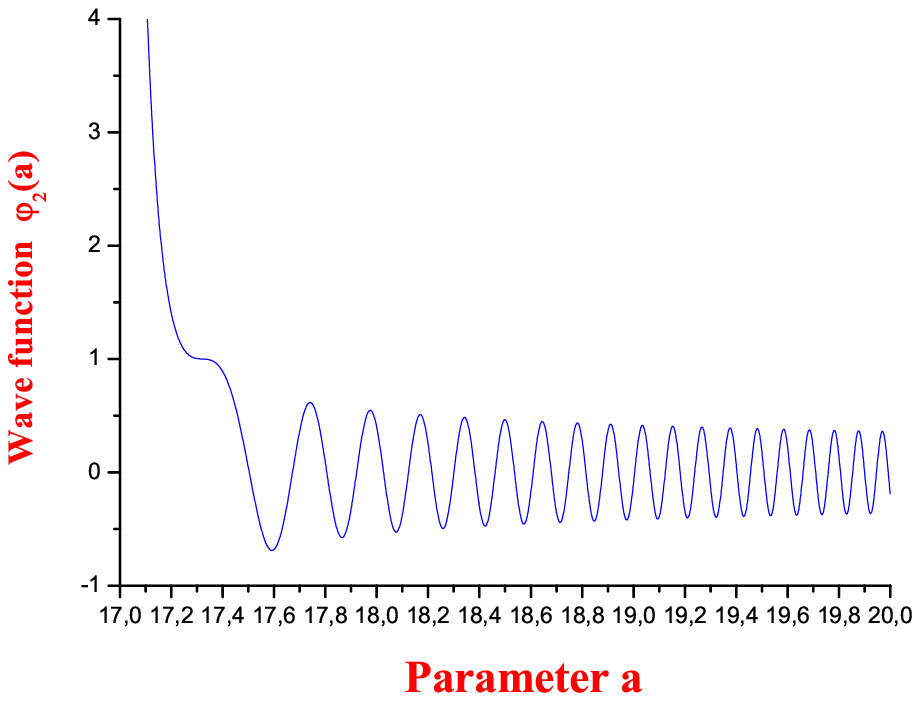}
\includegraphics[width=55mm]{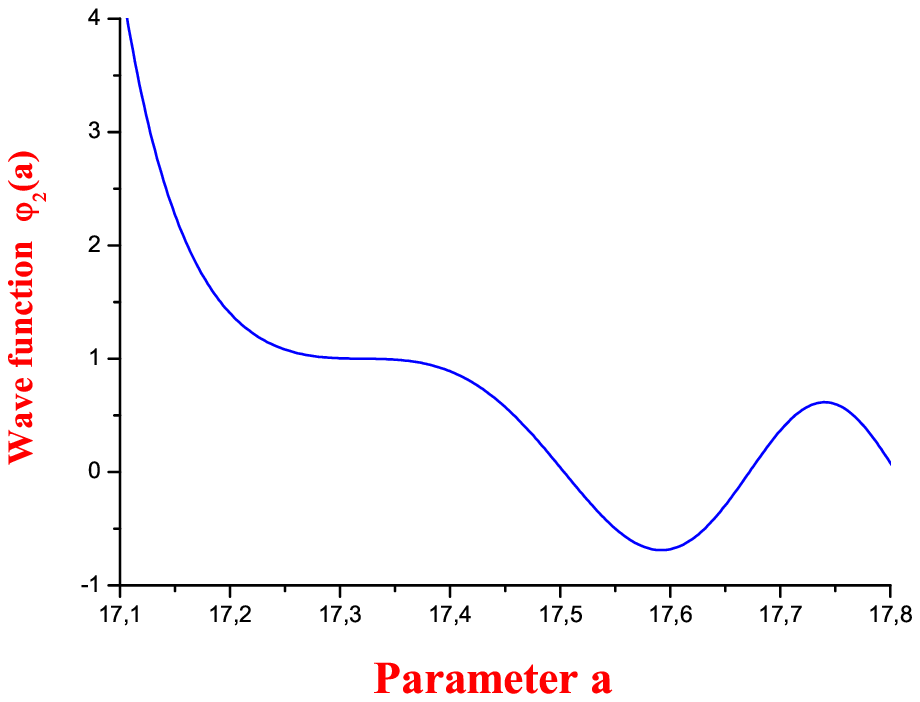}}
\caption{\small The behavior of the wave function in the selected region:
(a) the wave function in the tunneling region (for the vertical axis we use $\ln{(\varphi_{2}(a))}$);
(b) the wave function in the region of the escape point;
(c) the smooth transformation of the decrease of the wave function into the first oscillation close to the point $a_{tp}$.
\label{fig.6}}
\end{figure}

Let us analyze the behavior of the wave function more carefully. In Fig.~\ref{fig.6} its shape in the region of tunneling of the wave and close to the escape point is shown. In the first figure one can see its monotonous logarithmic decrease (without oscillations) in the tunneling region with increasing $a$. In the second figure the behavior of the wave function in the region of the escape point is shown, where the first oscillations appear. In the third figure one can see how a smooth decrease of the wave function is transformed little by little into the first oscillation in the small neighborhood of point $a_{tp}$, and the boundary of such a transformation must be the escape point $a_{tp}$ (as in the first solution, here the rules of behavior for the wave function in the sub-barrier and above barrier regions are shown too).
As we see, the proposed approach gives smooth pictures without divergences both for the wave function and its derivative.

Analyzing all pictures, we point out the following properties.

\begin{itemize}

\item
The principally different behavior of the two found partial solutions for the wave function (and its derivative) in the tunneling region ensures their linear independence. This allows one to construct a stable general solution for the wave function and for its derivative on their basis.

\item
A property of \emph{the uniform decrease of the oscillation period in the above barrier region for the wave function and its derivative with increasing $a$}, found for the first solution (increasing in the tunneling region), exists for the second solution (decreasing in the tunneling region) as well.

\item
\emph{As this property is satisfied for each linearly independent partial solution, it must be satisfied for the total wave function and its derivative as well. From this one can obtain new information on the specific character of the dynamics of the Universe expansion (and to use the oscillation period as a new characteristic for the description (estimation) of this dynamics).}

\item
Two partial solutions for the wave function and its derivative and their analysis are obtained at zero energy.
\end{itemize}

\subsection{Determination of the total wave function in the whole semiaxis of $a$
\label{sec.3.3}}

After obtaining two linearly independent partial solutions $\varphi_{1}(a)$ and $\varphi_{2}(a)$ for the wave function at the potential (\ref{eq.intro.24}), we write the general solution as
\begin{equation}
  \varphi (a) = N \cdot \bigl(C_{1}\, \varphi_{1}(a) + C_{2}\,\varphi_{2}(a) \bigr),
\label{eq.3.3.1}
\end{equation}
where $C_{1}$ and $C_{2}$ are arbitrary complex constants, $N$ is a normalization factor (we separate it explicitly from the constants $C_{1}$ and $C_{2}$, because these constants and the factor $N$ are found from different conditions, and they have different effects on the penetrability coefficient and other characteristics).

One can find the unknown constants $C_{1}$ and $C_{2}$ from the tunneling boundary condition pointed out above:
\emph{the function $\varphi(a)$ must define only an outgoing wave at the escape point $a_{tp}$}. Here, the function $\varphi(a)$ and its derivative must be equal to (\ref{eq.3.1.19}) at the point $a_{tp}$ (we shall use the factor $N$ for $\varphi(a)$ in consistence with the factor $N$ in (\ref{eq.3.1.19})):
\begin{equation}
\begin{array}{cc}
  \varphi(a_{tp}) = N \Psi (\xi=0), &
  \hspace{5mm}
  \displaystyle\frac{d \varphi(a)}{da} \bigg|_{a=a_{tp}} = N \displaystyle\frac{d \Psi (\xi)}{da} \bigg|_{a=a_{tp}} =
  N |V_{1}|^{1/3} \cdot \displaystyle\frac{d \Psi (\xi)}{d\xi} \bigg|_{\xi=0}.
\end{array}
\label{eq.3.3.2}
\end{equation}
Taking into account eq.~(\ref{eq.3.3.1}), we find (a prime is for a derivative with respect to $a$)
\begin{equation}
\begin{array}{cc}
  C_{1} = \displaystyle\frac{\Psi\varphi_{2}^{\prime} - \Psi^{\prime}\varphi_{2}}
          {\varphi_{1}\varphi_{2}^{\prime} - \varphi_{1}^{\prime}\varphi_{2}} \bigg|_{a=a_{tp}}, &
  C_{2} = \displaystyle\frac{\Psi^{\prime}\varphi_{1} - \Psi\varphi_{1}^{\prime}}
          {\varphi_{1}\varphi_{2}^{\prime} - \varphi_{1}^{\prime}\varphi_{2}} \bigg|_{a=a_{tp}}.
\end{array}
\label{eq.3.3.3}
\end{equation}

\begin{figure}[h]
\centerline{
\includegraphics[width=75mm]{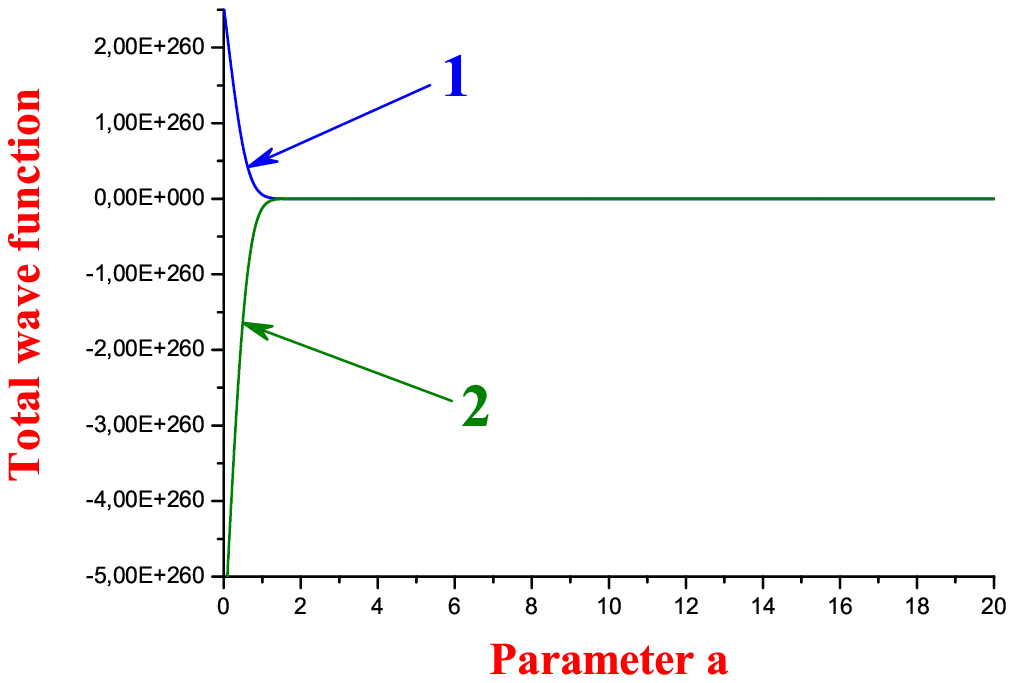}
\includegraphics[width=75mm]{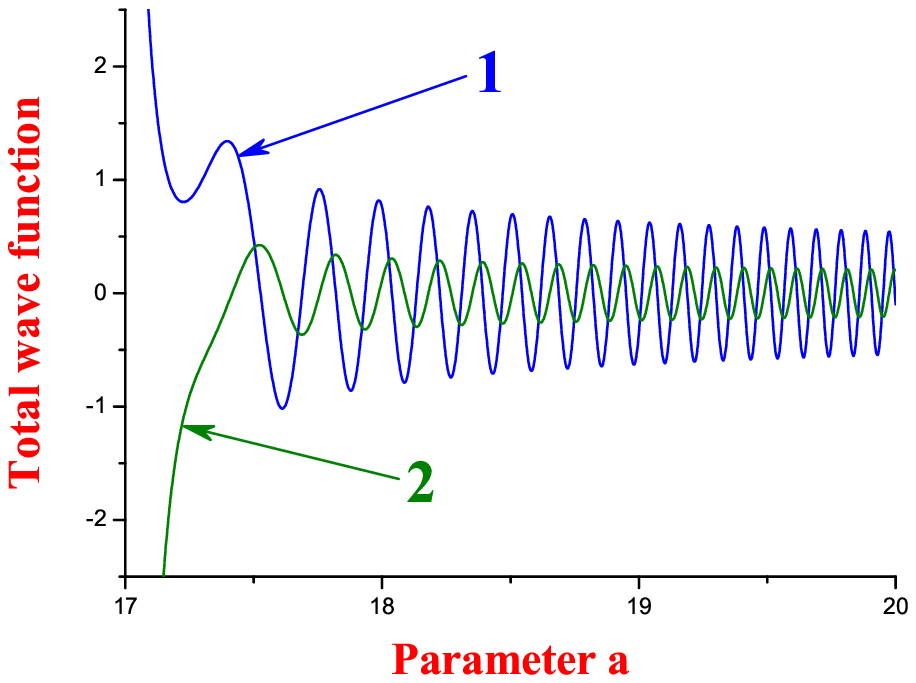}}
\caption{\small
The total wave function: curve 1, blue, is for the real part; curve 2, green, for the imaginary part.
\label{fig.7}}
\end{figure}

\begin{figure}[h]
\centerline{
\includegraphics[width=55mm]{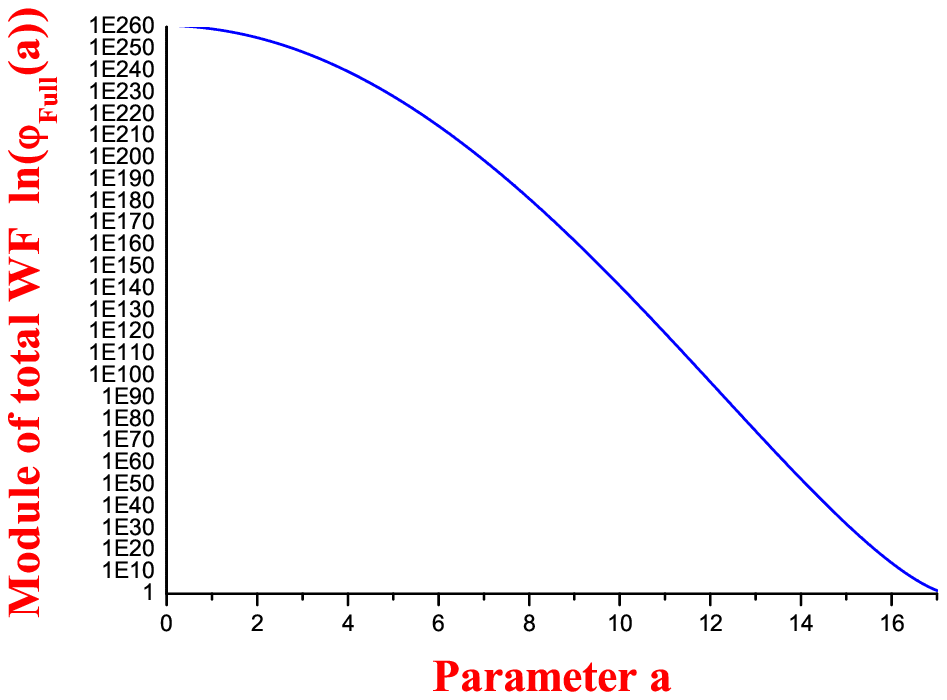}
\includegraphics[width=55mm]{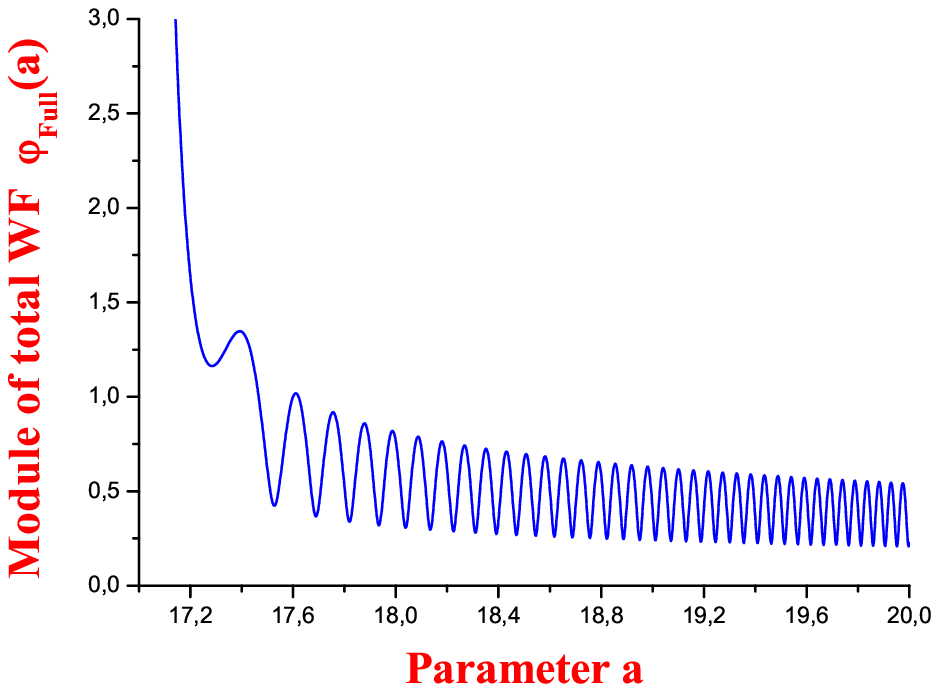}
\includegraphics[width=55mm]{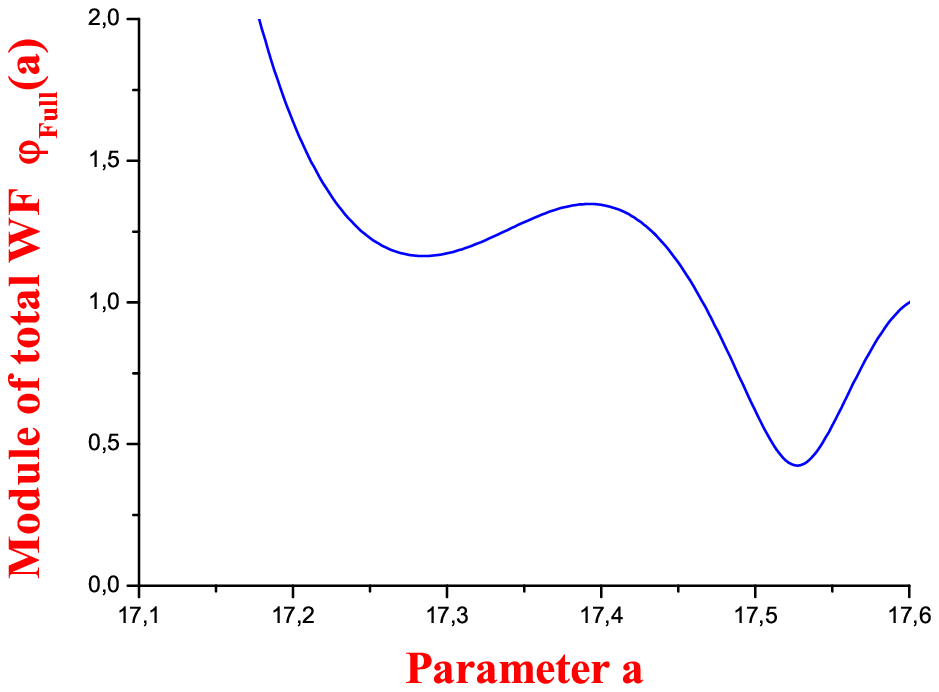}}
\caption{\small
The modulus of the total wave function:
(a) in the tunneling region the modulus decreases uniformly as a function of $a$;
(b) in the escape region, the minima and maxima appear in the wave function, which is connected with the existence of oscillations in it, but the modulus is never equal to zero in the whole range of $a$ (this demonstrates the existence of a \underline{non-zero} flux);
(c) near the escape point $a_{tp}$ we see that the modulus is changed minimally under increasing of $a$ (this demonstrates the fulfillment of the definition introduced above for the wave at point $a_{tp}$).
\label{fig.8}}
\end{figure}

The total wave function for the potential with the parameters from (\ref{eq.intro.25}), calculated by our approach, is shown in Fig.~\ref{fig.7}.
From Fig.~\ref{fig.8} one can analyze the behavior of the modulus of this wave function and from Fig.~\ref{fig.9} --- the behavior of its phase.
\begin{figure}[h]
\centerline{
\includegraphics[width=55mm]{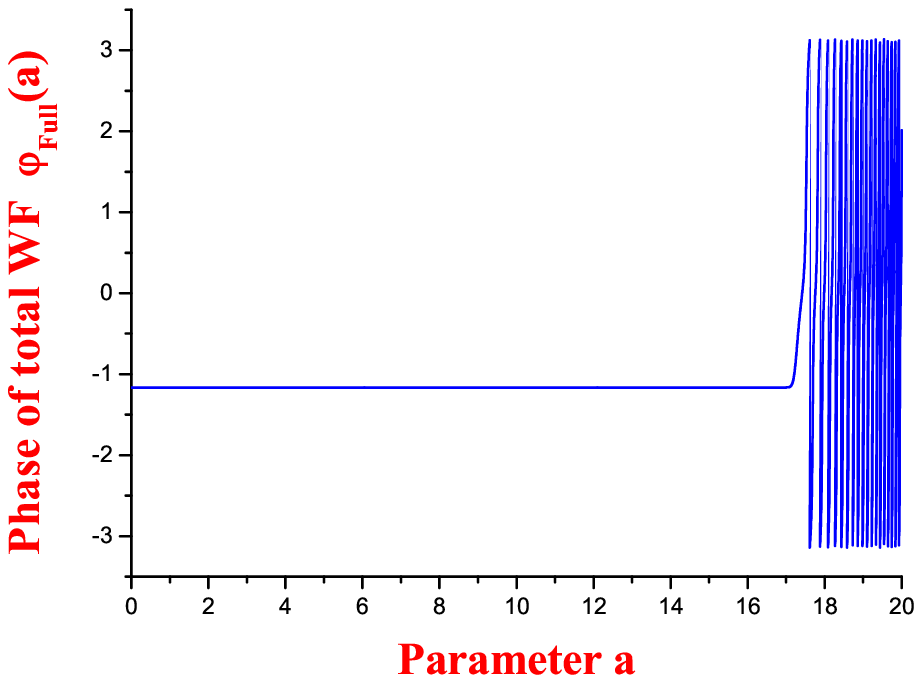}
\includegraphics[width=55mm]{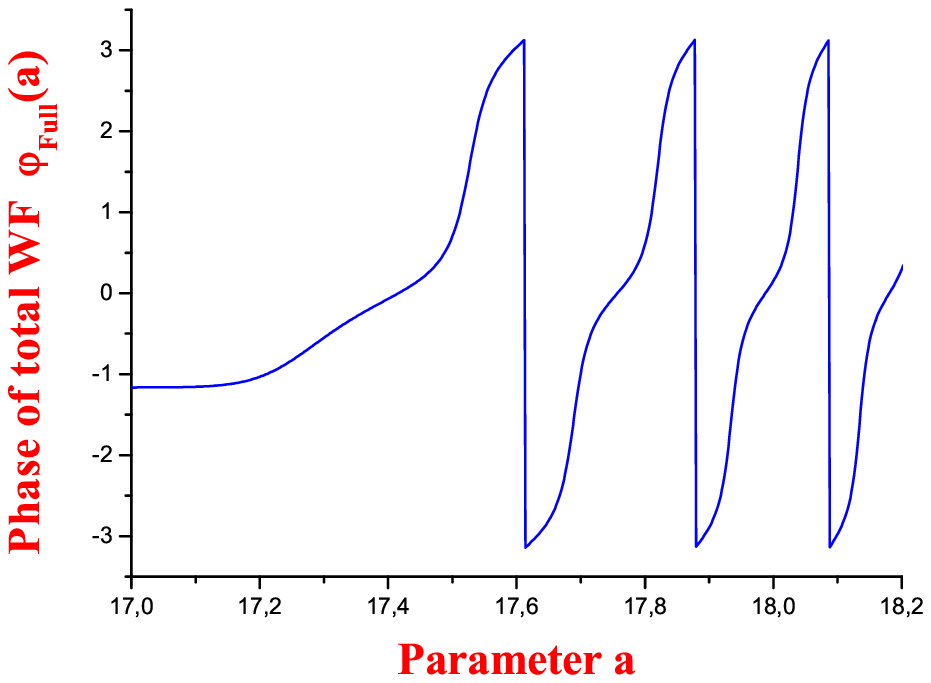}
\includegraphics[width=55mm]{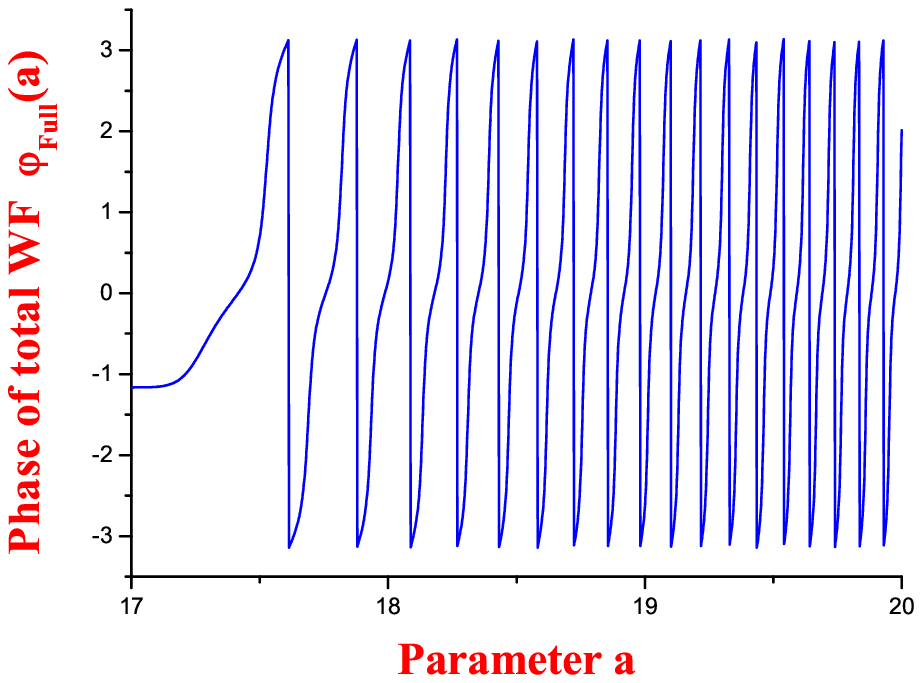}}
\caption{\small
The phase of the total wave function:
(a) in the barrier region the phase is constant, outside the barrier oscillations appear in the limits from $-\pi$ to $\pi$;
(b) the smooth appearance of the first oscillations of the phase in the neighborhood of the escape point $a_{tp}$ is shown;
(c) the oscillation period of the phase decreases uniformly with increasing $a$.
\label{fig.9}}
\end{figure}

\section{Estimation of the barrier penetrability
\label{sec.4}}

For the determination of the barrier penetrability in standard quantum mechanics for the wave escaping from this barrier outside we must know the incident flow inside the internal region. This flow must be constructed on the basis of the wave already tunneling under the barrier starting from point $a=0$ to the right. This method requires an accurate analysis and some questions may appear. Therefore, with the purpose to estimate the penetrability of the barrier studied, we restrict ourselves only to two other characteristics (which are analyzed often in the development of quantum cosmological models):
\emph{we define the probability $P_{\rm tun}$ of the appearance of the particle studied inside the tunneling region and the probability $P_{\rm ext}$ of the absence of it as the corresponding ratios of the probability of the wave functions, defined on the tunneling region and on the external region, to the probability of the wave function, defined over the whole range of its definition}:
%
%
\begin{equation}
\begin{array}{cc}
  P_{\rm ext} = \displaystyle\frac{\int\limits_{a_{tp}}^{a_{\rm max}} |\varphi(a)|^{2} \; da}
                        {\int\limits_{0}^{a_{\rm max}} |\varphi(a)|^{2} \; da}, & \hspace{10mm}
  P_{\rm tun} = \displaystyle\frac{\int\limits_{0}^{a_{tp}} |\varphi(a)|^{2} \; da}
                        {\int\limits_{0}^{a_{\rm max}} |\varphi(a)|^{2} \; da}
\end{array}
\label{eq.4.1.1}
\end{equation}
where $a_{\rm max}$ is the upper limit of the range of $a$ where we consider the total wave function $\varphi(a)$. Note that we use the stationary wave function $\varphi(a)$ at zero energy in contrast to~\cite{AcacioDeBarros.2006}. In order to keep the computer calculations as accurate as possible, we calculate $P_{\rm ext}$ thus:
\begin{equation}
  P_{\rm ext} = 1 - P_{\rm tun}.
\label{eq.4.1.2}
\end{equation}
The results of our calculations of $P_{\rm tun}$ are presented in Table~\ref{table.1}.
\begin{table}
\caption{Probability $P_{\rm tun}$ in dependence on the upper limit $a_{\rm max}$ and decomposition of the whole range of $a$.
\label{table.1}}
\hspace{-20mm}
\begin{center}
\begin{tabular}{|c|c|c|c|c|c|c|} \hline
 $a_{\rm max}$ &  \multicolumn{6}{|c|}{Decomposition} \\ \cline{2-7}
          &
    10000 &
    50000 &
   100000 &
   200000 &
  1000000 &
  2000000 \\ \hline
    20 &  0.0428858066  &  0.0420408533  &  0.0418943250  &  0.0419056013  &  0.0418471293  &  0.0418398319  \\
    30 &  0.0230343856  &  0.0229506126  &  0.0230091568  &  0.0229690161  &  0.0229369614  &  0.0229329589  \\
    40 &  0.0197653508  &  0.0197616584  &  0.0195350357  &  0.0195796586  &  0.0195208513  &  0.0195095816  \\
    50 &        -       &  0.0183045342  &  0.0180612516  &  0.0180298553  &  0.0180046869  &  0.0180015439  \\
    70 &        -       &        -       &  0.0167357966  &  0.0165972606  &  0.0165979553  &  0.0165950583  \\
   100 &        -       &        -       &       -        &  0.0156822095  &  0.0156892697  &  0.0156865629  \\
   150 &        -       &        -       &       -        &        -       &  0.015056842   &  0.0150574352  \\
   200 &        -       &        -       &       -        &        -       &        -       &  0.0147582032  \\ \hline
\end{tabular}
\end{center}
\end{table}
One can see that increasing of the decomposition in the selected region gives a stable convergent value for $P_{\rm tun}$. This demonstrates the convergence of the proposed method in the calculation of the wave function. From the data we select the following limit:
\begin{equation}
\begin{array}{cc}
  P_{\rm tun} = 0.014, & \hspace{10mm}
  P_{\rm ext} = 1 - 0.014 = 9.986,
\end{array}
\label{eq.4.1.3}
\end{equation}
which differs significantly from the results of~\cite{AcacioDeBarros.2006}.

\subsection{Comparison with semiclassical calculations
\label{sec.4.2}}

Now we shall estimate the penetrability in the semiclassical approximation. First, let us consider the wave function in the case $E \ne 0$. Here, we have internal and external turning points, which we denote $a_{tp}^{\rm (int)}$ and $a_{tp}^{\rm (ext)}$. Let the semiclassical wave function in the region to the right from the external turning point $a_{tr}^{\rm (ext)}$ be a wave outgoing from the barrier outside having the form (we define it in the second approximation)
\begin{equation}
  \varphi_{\rm out}(a) =
  N \cdot f_{2}(a) \cdot
  \exp
  \Biggl\{
    i\,\displaystyle\int\limits_{a_{tp}^{\rm (ext)}}^{a} p\; da + \displaystyle\frac{i\pi}{4}
  \Biggr\},
\label{eq.4.2.1}
\end{equation}
where $f_{2}(a)=1 / \sqrt{|p\,(a)|}$, $p\,(a) = \sqrt{E - V(a)}$ is the complex momentum, and $N$ is a normalization factor. Then according to the rules of correspondence (50,2) and (47,5) in \cite{Landau.v3.1989} between wave functions in different regions close to the turning points, we have obtained the wave function $\varphi_{\rm tun}(a)$ in the barrier region and the wave function $\varphi_{\rm int}(a)$ in the left region before the barrier:
\begin{equation}
\begin{array}{cc}
  \varphi_{\rm tun}(a) =
  N \cdot f_{2}(a) \cdot
  \exp
  \Biggl| \displaystyle\int\limits_{a}^{a_{tp}^{\rm (ext)}} p\; da \Biggr| =
  N \cdot f_{2}(a) \cdot
  \exp
  \Biggl\{
    \biggl| \displaystyle\int\limits_{a_{tp}^{\rm (int)}}^{a_{tp}^{\rm (ext)}} p\; da \biggl| -
    \biggl| \displaystyle\int\limits_{a_{tp}^{\rm (int)}}^{a} p\; da \biggr| \Biggr\}, \\

  \varphi_{\rm int}(a) =
  2\, N \cdot f_{2}(a) \cdot
  \exp
  \Biggl( \displaystyle\int\limits_{a_{tp}^{\rm (int)}}^{a_{tp}^{\rm (ext)}} |p|\; da \Biggr) \cdot
  \cos
  \Biggl\{
    \displaystyle\int\limits_{a}^{a_{tp}^{\rm (int)}} p\; da -
    \displaystyle\frac{\pi}{4}
  \Biggr\}.
\end{array}
\label{eq.4.2.2}
\end{equation}

In Fig.~\ref{fig.9} one can see the modulus of the wave functions calculated in the first and second semiclassical approximations (at $E=0$) located together with the modulus of the wave function obtained by the direct quantum method proposed above. Here, we calculate the wave function in the second semiclassical approximation inside the barrier region by the first formula in~(\ref{eq.4.2.2}) and in the external region by the second formula in~(\ref{eq.4.2.2}). The wave function in the first semiclassical approximation we calculate by the same formulas and use $f_{2} = 1$. We normalize both solutions found separately on the wave function obtained by the direct approach, using the factor $N$.
\begin{figure}[h]
\centerline{
\includegraphics[width=55mm]{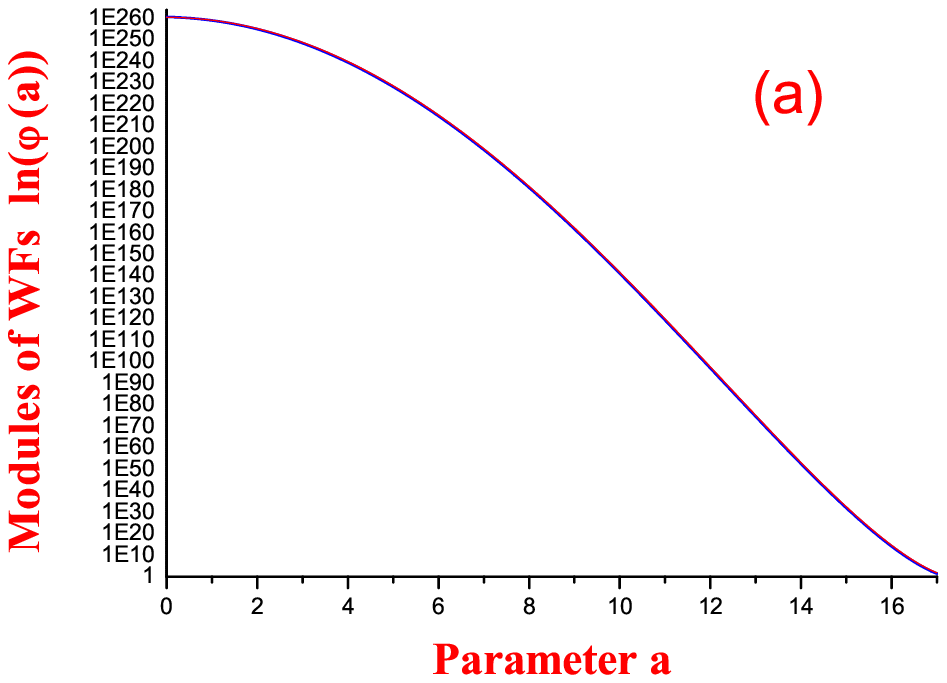}
\includegraphics[width=55mm]{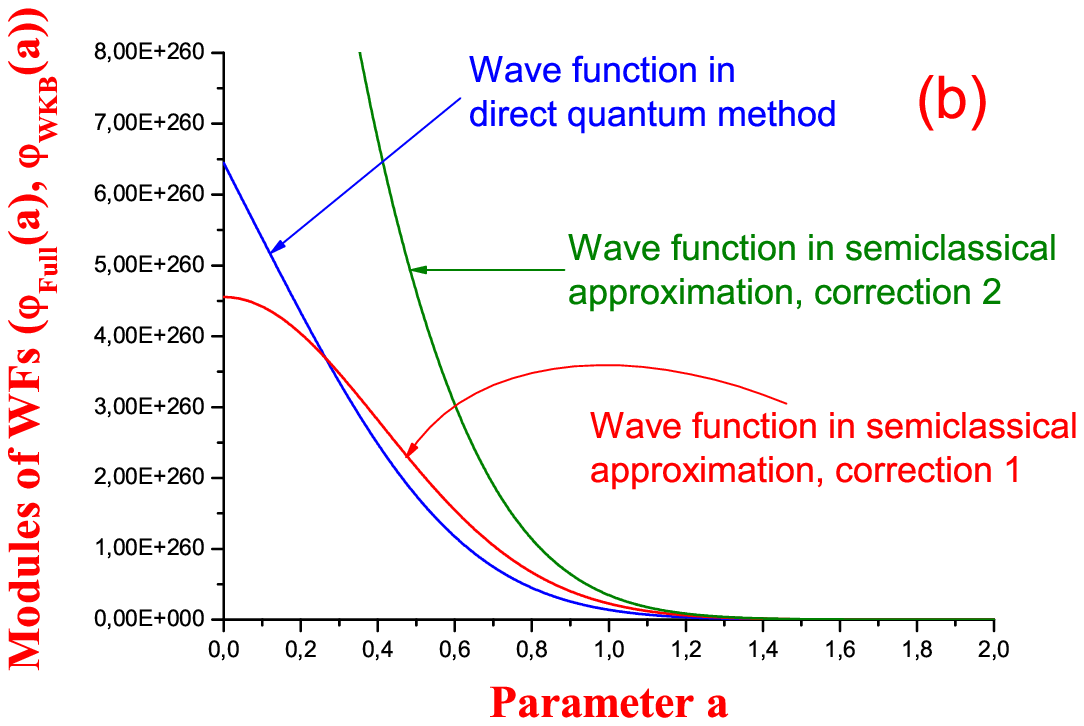}
\includegraphics[width=55mm]{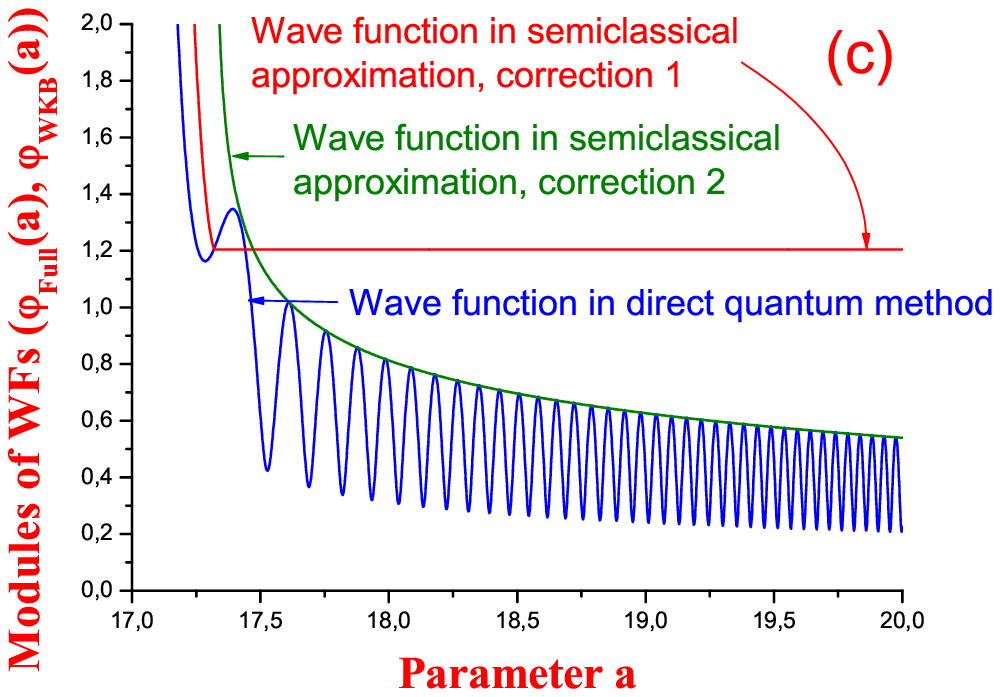}}
\caption{\small
The modulus of the wave function calculated by the direct quantum approach and wave functions calculated in the first and second semiclassical approximations:
(a) in the consideration inside the whole barrier region, we obtain a complete and very precise coincidence between the modulus of the wave function calculated by the direct approach and both wave functions calculated by two semiclassical approximations (except for a region close to zero for the solution in the second WKB approximation, which singularity is not visible at such scale);
(b) close to zero the wave function in the first WKB approximation describes more accurately the modulus of the wave function in the direct approach in comparison with the wave function in the second WKB approximation (at $a=0$, the wave function in the second WKB approximation has a singularity, while we obtain finite values for the wave function in the direct approach and the wave function in the first WKB approximation);
(c) near the escape point $a_{tp}$ the modulus of the wave function in the second semiclassical approximation coincides with the maxima of the wave function in the direct approach.
\label{fig.9}}
\end{figure}
From these figures we conclude to the following.
\begin{itemize}

\item
Inside the whole barrier region the modulus of the wave function in the direct approach is approximated very precisely and accurately by the wave functions in the first and second approximation. Close to zero it is described better by the wave function in the first approximation: for the wave functions in the second approximation we have a singularity while for the wave function in the direct approach and for the wave function in the first approximation we have obtained finite values at point $a=0$. In a detailed study of three solutions we observe their values to be different (with a similar tendency), which can be shown in further calculations of the penetrability and similar characteristics. By this, one can explain different possible (estimated) values of the penetrability at the very precise coincidence between the exact and semiclassical solutions presented.

\item
Inside the external region the wave function in the first approximation can be considered as a ``plane'' wave (with constant modulus and changed phase). We see that it describes less accurately the exact solution while the wave function in the second approximation coincides very accurately with maxima of the wave function in the direct approach, and it can be considered as a proper solution.
We observe the presence of oscillations of the modulus of the wave function in the direct approach which decrease with increasing of $a$ (and which are not present in semiclassical solutions).
Such oscillations may indicate the following supposition:
\emph{the outgoing wave in the external region close to the turning point cannot be considered as a usual ``plane'' wave, and oscillations include information of the interaction between the studied outgoing wave and the barrier at not too large distances.}
One can connect them with the phenomenon of the barrier non-locality:
the barrier has influence on the tunneling and further propagating wave as a unified object, and in estimating of the penetrability we consider the barrier as the potential in the whole region.
The rules of the correspondence between two semiclassical solutions of the wave function in different regions can break down (or change) the non-locality of the barrier near the turning point, and we obtain a greater local influence of the barrier on the semiclassical solutions of the wave function in different regions separately.

\item
Close to the turning point the wave function in the direct approach connects both semiclassical solutions for the wave function in tunneling and external regions. Such a connection can be considered as a small shift of the wave function along the $a$ axis (perhaps this has been found for the first time in cosmological models).
\end{itemize}

Now on the basis of (\ref{eq.4.2.1}) and (\ref{eq.4.2.2}) we find the known formula for the penetrability coefficient in the semiclassical approximation:
\begin{equation}
  P_{\rm penetrability}^{\rm WKB} =
  \exp
  \Biggl(
    - 2\: \displaystyle\int\limits_{a_{tp}^{\rm (int)}}^{a_{tp}^{\rm (ext)}}
    |p|\; da \Biggr).
\label{eq.4.2.3}
\end{equation}
Taking into account only the first semiclassical approximation and comparing the first expression in (\ref{eq.4.2.2}) with (\ref{eq.4.2.3}) (at the needed normalization of $N$), we obtain a new formula:
\begin{equation}
  P_{\rm penetrability}^{\rm WKB,\, (1)} \sim
  \displaystyle\frac
    {\Bigl| \varphi_{\rm tun}(a_{tp}^{\rm (int)}) \Bigr|^{2}}
    {\int\limits_{0}^{a_{\rm max}} |\varphi(a)|^{2} \; da}.
\label{eq.4.2.4}
\end{equation}
%
Calculations in the semiclassical approximation gives (decomposition of the barrier region is in 100000 intervals):
\begin{equation}
\begin{array}{lcl}
  P_{\rm tun}^{\rm WKB} = 0.052, & \hspace{10mm}
  P_{\rm penetrability}^{\rm WKB} = e^{-1200} = 7.0 \cdot 10^{-522}.
\end{array}
\label{eq.4.2.5}
\end{equation}
Now we see that the direct approach gives the $P_{\rm tun}^{\rm direct}$ coefficient 4--5 times smaller than $P_{\rm tun}^{\rm WKB}$. This points to the existence of more essential tunneling processes inside the barrier. One may also suppose that taking the external region into account we can change the value of the penetrability coefficient $P_{\rm penetrability}^{\rm WKB}$ found.


\section{Conclusions and perspectives
\label{sec.conclusions}}

In this paper a new method of (non-semiclassical) calculation of the wave function of the Universe in the quantum cosmological model of a Universe of closed type, in the framework of the Friedmann--Robertson--Walker metric, is presented. We note the following.
\begin{itemize}
\item
A method for the calculation of two partial solutions for the wave function and its derivative is constructed:

\begin{itemize}
\item
at first, the values for the wave function and its derivative are defined at a starting point (for the partial solution which increases in the barrier region, we use the starting point $a=0$, and for the second partial solution, which decreases in the barrier region, we select the starting point to be the escape point $a_{tp}$);

\item
using the algorithms in Secs.~\ref{sec.3.2.1} and \ref{sec.3.2.2}, both partial solutions are found independently in the vicinity of the starting point;

\item
using the Numerov method with a constant step, both partial solutions are found in the whole studied range of $a$.
\end{itemize}

\noindent
In this way, we obtain a well-convergent picture for the wave function and its derivative (for the potential from Ref.~\cite{AcacioDeBarros.2006}, the convergence of the solutions is demonstrated up to $a<100$).

\item
In order to describe the process of tunneling as accurately as a possible, to construct the total wave function on the basis of its two partial solutions unambiguously, we use the tunneling boundary condition from papers of A.~Vilenkin, at the escape point $a_{tp}$:
\emph{the total wave function must represent only the outgoing wave}.

\item
We introduce two definitions of this wave in the (right) neighborhood of the boundary $a_{tp}$ as follows.

\begin{itemize}
\item
The wave is such a linear combination of two partial solutions of the wave function that a change the modulus $\rho$ of this wave function is the closest to constant under variation of $a$ (strict definition of the wave; see (\ref{eq.3.1.8}))

\item
The wave is such a linear combination of two partial solutions of the wave function that the modulus $\rho$ changes minimally under variation of $a$ (weak definition of the wave; see (\ref{eq.3.1.9})).
\end{itemize}

\item
Such definitions of the wave allow us to construct sufficiently stable and convergent solutions for the total wave function and its derivative (for the potential from~\cite{AcacioDeBarros.2006} the solutions are calculated at $a<100$).

\item
Analyzing the wave function, the following properties have been shown.

\begin{itemize}
\item
\emph{We observe oscillations of the modulus of the wave function in the external region starting from the turning point $a_{tp}$ which decrease with increasing of $a$, and these are not shown in semiclassical calculations} (perhaps, they have been found for the first time). An interesting idea is in the physical interpretation: their maxima an minima point to values of $a$ where the modulus of the wave function of the Universe is maximal or minimal, and one may suppose that such an effect shows specific peculiarities of space-time of the Universe in smaller distances while it disappears at large ones, tending to the semiclassical limit.

\item
\emph{With increasing $a$, the period of oscillations decreases uniformly smoothly both for the wave function, and for its derivative in the above barrier region.} This peculiarity is fulfilled for each partial solution, and, therefore, it must appear in the total wave function and its derivative as well. From this one can find new information on the specific character of dynamics of the expansion of the Universe (and one can use the oscillation period as a certain characteristic for the estimation of the dynamics of such an expansion).

\begin{itemize}
\item
This explains why for a sufficiently small increase of $a$ it becomes \underline{significantly more difficult} to calculate the convergent continuous solutions for the wave function and its derivative (where it is required to reduce the step significantly)!

\item
This information may be interesting for the choice or construction of new functions, by which the total wave function in the region of interest can be expanded with the highest efficiency (for example, this is why searching for solutions for the total wave function as an expansion in plane waves or spherical Bessel functions cannot give a sufficiently stable convergent result (with a small increase of the selected range of $a$)).

\item
In order to calculate the wave function and its derivative with higher efficiency, we propose, instead of the Numerov method with a the constant step, to use another method of continuation of the solution, where the step decreases uniformly with increasing $a$.
\end{itemize}
\end{itemize}


\end{itemize}

\section*{Acknowledgements
\label{sec.acknowledgements}}

\begin{acknowledgement}
The author appreciates 
useful discussions with Dr.~Volodymyr Uleshchenko concerning the application of quantum non-semiclassical approaches in quantum cosmology and determination of wave functions of the Universe inside the tunneling region,
and he is graceful to
Dr.~Yevgeny~Kats for his assistance in preparing the paper,
and Dr.~Mariam Bouhmadi Lopez for useful comments concerning different aspect of quantum cosmology of asymptotically de Sitter Universes.
\end{acknowledgement}

\bibliography{Cosmology_epjc_v3}

\end{document}